\documentclass[11pt,a4paper]{article}
\usepackage{amsmath}
\usepackage{dsfont}
\usepackage{bm}
\usepackage{accents}
\usepackage{amsbsy}

\usepackage{a4wide,graphicx,times,psfrag,wrapfig,sidecap}
\usepackage{cite}
\usepackage[colorlinks=true,linkcolor=black, citecolor=black,
urlcolor=black]{hyperref}
\numberwithin{equation}{section}
\makeatletter \let\old@startsection=\@startsection
\renewcommand{\@startsection}[6]
{\old@startsection{#1}{#2}{#3}{#4}{#5}{#6\mathversion{bold}}}
\makeatother

\def\<{\langle}
\def\>{\rangle}

\def\tr{{\rm   tr} }
\def\Im{{\rm Im}}

\newcommand\encadremath[1]{\vbox{\hrule\hbox{\vrule\kern8pt
\vbox{\kern8pt \hbox{$\displaystyle #1$}\kern8pt}
\kern8pt\vrule}\hrule}} \def\enca#1{\vbox{\hrule\hbox{
\vrule\kern8pt\vbox{\kern8pt \hbox{$\displaystyle #1$} \kern8pt}
\kern8pt\vrule}\hrule}}
  \usepackage{bm}

\def\Xint#1{\mathchoice
   {\XXint\displaystyle\textstyle{#1}}%
   {\XXint\textstyle\scriptstyle{#1}}%
   {\XXint\scriptstyle\scriptscriptstyle{#1}}%
   {\XXint\scriptscriptstyle\scriptscriptstyle{#1}}%
   \!\int}
\def\XXint#1#2#3{{\setbox0=\hbox{$#1{#2#3}{\int}$}
     \vcenter{\hbox{$#2#3$}}\kern-.5\wd0}}

\def\dashint{\Xint-}

\usepackage{amssymb}
\usepackage{soul}
\makeatletter
\renewcommand*\env@matrix[1][\arraystretch]{%
  \edef\arraystretch{#1}%
  \hskip -\arraycolsep
  \let\@ifnextchar\new@ifnextchar
  \array{*\c@MaxMatrixCols c}}
\makeatother
\begin{document}

\thispagestyle{empty}
\vspace{1cm}
\setcounter{footnote}{0}
\begin{center}
{\Large \ Entanglement   Spectrum in General Free Fermionic Systems}

\vspace{20mm}
Eldad Bettelheim$^{1,2}$, Aditya Banerjee$^1$, Martin B. Plenio$^2$, Susana F. Huelga$^2$   \\[7mm]

$^1$:Racah Institute of Physics, Hebrew University of Jerusalem, Edmund J Safta Campus\\ 91904 Jerusalem, Israel\\
$^2$: Institute of Theoretical Physics and Center for Integrated Quantum Science and Technology (IQST), Universit\"{a}t Ulm, Albert-Einstein-Allee 11, Ulm 89069,
Germany.
\vspace{20mm}

\end{center}

\vskip9mm

\vskip18mm

\noindent{ The statistical mechanics characterization  of a finite subsystem embedded in an infinite system is a fundamental question of quantum physics. Nevertheless, a full closed form  { for all required entropic measures} does not exist in the general case even for free systems when the finite system in question is composed of several disjoint intervals. Here we develop a mathematical framework based on the Riemann-Hilbert approach to treat this  problem in the one-dimensional case where the finite system is composed of two disjoint intervals and in the thermodynamic limit (both intervals and the space between them contains an infinite number of lattice sites and the result is given as a thermodynamic expansion).  To demonstrate the usefulness of our method, we compute the change in the entanglement and negativity namely the spectrum of eigenvalues of the reduced density matrix with our without time reversal of one of the intervals. We do this in the case that the distance between the intervals is much larger than their size. The method we use can be easily applied to compute any power in an expansion in the ratio of the distance between the intervals to their size.  {We expect these results to provide the necessary mathematical apparatus to address relevant questions in concrete physical scenarios, namely the structure and extent of quantum correlations in fermionic systems subject to local environment. }       }

\section{Introduction}
A basic concept in statistical physics is the thermodynamics of a subsystem of a larger, often infinite system, the most basic example being perhaps the derivation of the  canonical ensemble from the micro-canonical one by taking an infinite system in the micro-canonical ensemble and considering a smaller sub-system, which is then found to be in the canonical ensemble. Nowadays, when applying this procedure to quantum systems and when the thermodynamic potential being investigated in the small system is the entropy, the quantity being measured is brought under the title of 'entanglement entropy'. 

Although the newfangled `entanglement entropy' is, in fact, the mundane `entropy' of a sub-region in newer attire, the realization that it is (also) a measure of entanglement allows to focus on aspects previously ignored and for interesting generalizations of the basic quantity \cite{Eisert:Cramer:Plenio:Colloquium}. For example, the entropy of the union of several sub-regions may be explored and how it differs from the sum of the individual entropies of the regions may be considered. Another example, comes under the name of 'logarithmic negativity'\cite{plenio:Logarithmic:Negativity:Monotone}. For the case of two regions it  involves reversing the arrow of time on one sub-region while keeping its direction for the other sub-region and then computing a quantity which is obtained from the entanglement spectrum (to be defined below) of the union of the two sub-regions. The resulting quantity is a measure of entanglement which also is valid for mixed states (in contrast to the entanglement entropy) of the union of the two regions\cite{peres:separability,horodeckis:separability,eisert-plenio:Comparison,plenio:Logarithmic:Negativity:Monotone}.

This manuscript is concerned with the computation of the spectrum of eigenvalues of the reduced density matrix  which both in the case in which one applies a partial transpose operation as well as in the case it is not applied in a one dimensional translationally invariant open or closed system of free fermions, from which the logarithmic negativity and entanglement entropy can be computed. These have been computed in the past {using {numerical methods on bosonic lattice systems \cite{Marcovitch:Retzker:Plenio:Reznik:Critical:Noncritical:Long:Range:Entanglement:},}}  field-theoretic techniques\cite{Calabrese:Cardy:Tonni:Negativity:Field:Theoretic,Calabrese:Cardy:Tonni:Negativity:QFT,Casini:Huerta:Two:Blocks} and conformal field theory techniques \cite{Calabrese:Cardy:Tonni:Negativity:Two:Intervals:II,Calabrese:Cardy:Tonni:Negativity:Two:Intervals:I}. Closed form results are available when the two sub-systems  are adjacent for logarithmic negativity. We therefore investigate here the case where the two sub-systems are not adjacent. It should be noted that the reversal of the time arrow or transposition of the states in one of the regions for fermions has been worked out satisfactorily  only recently in Refs. \cite{shapourian:Hasan:Shiozako:ken:Ryu:Shinsei:Negativity,Shapourian:Ruggiero:Ryu:Twisted:Untwiste:Negativity}. These references also provide results for the different quantities, but again the two sub-regions are adjacent.

Our approach here is akin to that of Ref. \cite{Jin:Korepin:Entanglement:Entropy:Fermions}, where one starts with the observation that  correlation functions within a sub-region are trivially equal to those in the entire region, and that for a free Gaussian system the correlation functions completely dictate the free action that describes the sub-region\cite{Peschel:Correlation}. From the free Gaussian action of the sub-region, in turn, it is easy to compute the entropy of the sub-region.
Note that the Gaussian action of the entire region is necessarily different from {one} of the sub-region. In field-theoretic language the action of the sub-region is obtained by integrating out the degrees of freedom outside the sub-region.  The fact that the correlation function in the sub-region (irrespective of whether we treat it as a sub-system of the large system or as an independent system) obey Wick's theorem, has two important consequences. The first is that the sub-system must be described by a free Gaussian action, and the second  is that it is enough to study the two-point function. 

The two point function in a one dimensional  single interval can be thought of as a matrix with the indices enumerating the two points involved within the region. The matrix is a Toeplitz matrix since the elements of the matrix depend only on the distance between the two points for a translationally invariant system. Finding the eigenvalues of this matrix {provides} enough information in order to compute the entropy of that region. To accomplish the latter the authors of Ref. \cite{Jin:Korepin:Entanglement:Entropy:Fermions} applied the Fisher-Hartwig theorem\cite{fisher:hartwig:Theorem}, which gives the determinants of Toeplitz matrices when the size of the matrix is  large. A determinant predicts the eigenvalues of a matrix $A$ by the the usual method of characteristic polynomials. Indeed, one computes $\det(\lambda-A)$ and searches for the zeros of this determinant. Since for $A$\   a Toeplitz matrix  also  $ \lambda-A$ is Toeplitz, it is possible to apply the Fisher-Hartwig theorem to compute the characteristic polynomial of $A$, namely   $ \det(\lambda-A),$  which in turn gives the eigenvalues of $A$ .

The moment  one deals with more than one interval, however, the  Toeplitz property of the correlation function is lost. In addition, if one wants to compute the logarithmic negativity of a region, one must apply a certain  transformation on the covariance matrix, thereby destroying the Toeplitz property even for adjacent regions. To deal with this situation, we go back to how the Fisher-Hartwig theorem is proved, and try to apply the fundamental methods used to prove the Fisher-Hartwig theorem\cite{Deift:Its:Krasovksy:Toeplitz:Hankel,Deift:Its:Krasovsky:Toeplitz} on the covariance matrix of two regions directly (possibly deforming the covariance matrix appropriately  to compute logarithmic negativity), without using a ready-made theorem. The method involved\cite{Deift:Its:Krasovksy:Toeplitz:Hankel,Deift:Its:Krasovsky:Toeplitz,Its:Krasovsky:Gaussian:With:Kump} turns out to be the Riemann-Hilbert problem coupled with the orthogonal polynomial technique.  It turns out that these methods will allow us to obtain the desired result. Indeed, the Riemann-Hilbert problem associated with finding the determinant of a Toeplitz matrix (thus producing the Fisher-Hartwig theorem), is closely related to the Riemann-Hilbert problem that we shall solve here, as both are solved in terms of confluent hypergeometric functions\cite{Its:Krasovsky:Gaussian:With:Kump}. 

We dedicate this paper to the technical aspects of the computation, whereby we show how to {  map the problem of entanglement characterisation into a Riemann-Hilbert problem, 
 }  and plan to apply the method to obtain more physical results in a following publication. {Nevertheless, we give here some results as well, notably  Eqs. (\ref{finalfinal1}-\ref{finalfinal2}), the entanglement and negativity spectra to leading order in the ratio between the size of the intervals and their distance. }

\section{Open and Closed Free Fermion Systems on a Line}

\begin{figure}[h!!!]
\begin{center}
\includegraphics[width=15cm]{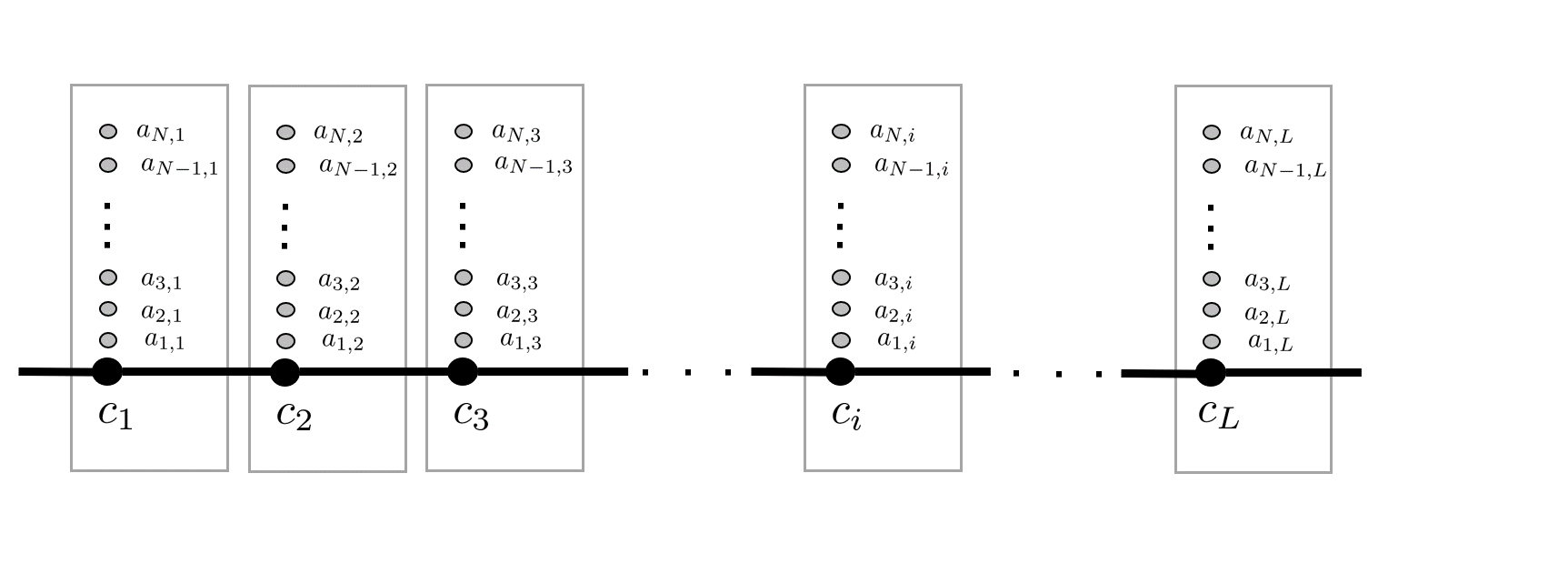}
\caption{
A system of Fermions described by the fermionic  annihilation operator $c_i$ of $L$ sites are coupled to each other by nearest neighbour interaction (heavy lines). In addition, each fermion at site $i$ is coupled locally to $N$\ fermionic modes $a_{i,n}, $ with $n$ running from $1$ to $N$. The gray rectangular boxes denote that the fermions  contained in a given box interact with each other.   \label{cartoon}} \end{center}
\end{figure}

 We consider a fermionic free field theory, described by a fermionic field operator $c_i$ at site $i$, { each} coupled to { a reservoir comprised of   $N$ fermionic modes. These modes are labeled by the fermionic  operator $a_{n,i}$ where $n$ labels the mode number $1\leq n \leq N$ and $i$ labels the site (see Fig. \ref{cartoon} for a pictorial representation of the physical situation). }  The Hamiltonian of the system is   given by
\begin{align}
&H=-\sum_{i} \frac{\hbar^2}{2Ma^2} c^\dagger_i (c_{i+1}+c_{i-1}-2c_i) +\alpha\sum_{i,n}  H_n \hbar\sqrt\Delta(   c^\dagger_i a_{n,i}+c_i a^\dagger_{n,i})+\\
&+\sum_{i,n}  \hbar G_na^\dagger_{n, i}a_{n,i},
\end{align}
where  \begin{align}G_n= \Delta n+\omega_{<}, \quad \hbar w_>= \hbar \Delta N+\hbar\omega_< ,\end{align} and $\omega_<$\ and $\omega_>$ are the band edges of the reservoir. The quantity $\Delta$ represents the level spacing of the Reservoir, and $H_n$ is a coupling constant, with an arbitrary dependence on $n$.  One obtains the following equations for the field variables:
\begin{align}
&\imath \hbar \partial_tc_i=\frac{\hbar^2}{2Ma^2}(c_{i+1}+c_{i-1}-2c_i)-\sum_n\alpha  a_{n,i}H_n \hbar\sqrt \Delta\ \\
&\imath \hbar \partial_t a_{n,i}=-a_{n,i} \hbar G_n-\alpha c_i H_n \hbar\sqrt \Delta 
\end{align}
Now Fourier transform both in time and in space according to the following transformation rule\begin{align}
g(z,\omega)=\frac{1}{2\pi}\sum_jz^{j} \int dt g_j (t)e^{-\imath \omega t}
\end{align}
yields the following equations of motion for the Fourier transformed field operators:
\begin{align}
&\left( \frac{z+z^{-1}-2}{2\hbar^{-2}Ma^2} +\hbar \omega\right)c(z,\omega) -\alpha\sum_n a_n(z,\omega)H_n\hbar\sqrt \Delta=0  \\ &\hbar \omega a_n  (z,\omega)-\hbar G_na_n  (z,\omega)-\alpha c(z,\omega) H_n\hbar\sqrt \Delta =0.\label{formaleq}
\end{align}
Thus,
\begin{align}
&\left( \frac{z+z^{-1}-2}{2\hbar^{-2}Ma^2} +\hbar \omega\right)c(z,\omega) -\alpha^2c(z,\omega)\sum_n\frac{ H_n^2 \hbar \Delta}{\omega-G_n} =0,
\end{align}
Thus we have the dispersion relation:\begin{align}
&\hbar \omega =\varepsilon(p)+\tilde \alpha(\omega), \label{DispersionRelation}  
\end{align}
where 
\begin{align}
&\varepsilon(p)=\frac{\hbar^2}{Ma^2}(1-\cos(p)),\quad z=e^{\imath p},\label{epsilonpiscos}\end{align}
and
\begin{align}
&\tilde\alpha(\omega)=\hbar \alpha ^2\Delta\sum_n\frac{ H_n^2 }{\omega-G_n}\label{tildealpha} .\end{align}The eigenvectors that leads to the eigenvalue $\hbar \omega$ are formally obtained by solving Eq. (\ref{formaleq}) to yield:
\begin{align}
a_n(z,\omega)=\frac{\alpha H_n \sqrt\Delta}{ \omega -G_n} c(z,\omega).
\end{align}
Thus  the combination
\begin{align}
\tilde c_\omega(z,t)=\frac{1}{C_\omega }\left(c(z,t)+ \sqrt\Delta \alpha\sum_n \frac{H_n a_{n}(z,t)}{ \omega -G_n}\right)
\end{align}
where 
\begin{align}
C_\omega = \sqrt{1+\Delta\alpha^2\sum_n \frac{H^2_n }{(\omega -G_n)^2}} \label{ComegaDef}
\end{align}
satisfies:
\begin{align}
&\nonumber\imath \hbar \partial_t \tilde c_\omega(z,t) =-\hbar \omega  \tilde c_\omega (z,t), \quad \{\tilde c_\omega(z,t),\tilde c^\dagger_\omega(z,t)\}=1.
\end{align}
Numbering all the solution of the dispersion for given $z$  by $\omega_j$, where $1\leq j \leq N+1$, we have:
\begin{align}
c(z,t)=\sum_j  \frac{\tilde c_{\omega_j}(z,t)}{C_{\omega_j}},
\end{align}
so that, after dropping the time variable due to the trivial time dependence, one obtains: 
\begin{align}
\<c^\dagger (z)c(z)\>=\sum_j \frac{\<\tilde c^\dagger_{\omega_j}(z)\tilde c_{\omega_j}(z)\>}{C^2_{\omega_j}}.\end{align}Now, in the ground state of the entire system, which includes both the reservoir and the fermionic system. All states, enumerated by $\omega_j(p),$ with energy $\hbar \omega_j(p)$ below a certain energy, called the Fermi energy and denoted by $\epsilon_F,$ are occupied $\<\tilde c^\dagger_{\omega_j}(z)\tilde c_{\omega_j}(z)\>=1 $ and all those above are unoccupied, $\<\tilde c^\dagger_{\omega_j}(z)\tilde c_{\omega_j}(z)\>=0, $ and as such we have: \begin{align}\<c^\dagger (z)c(z)\>=\sum_j\frac{\theta(\hbar \omega_j(p)<\epsilon_F)}{C^2_{\omega_j}}
\end{align}
From Eq. (\ref{DispersionRelation}) and (\ref{ComegaDef}) one can see
\begin{align}
C_\omega^{2}=\left|\frac{1}{\hbar}\frac{d\varepsilon(p)}{d\omega} \right|, 
\end{align}
then Eq. (\ref{epsilonpiscos}) yields:
\begin{align}
C_\omega^2=\frac{\hbar|\sin(p)|}{ma^2 |\omega'(p)|}
\end{align}
Such that:
\begin{align}
\<c^\dagger (p)c(p)\>=\sum_j\frac{|\hbar \omega_j'(p)|}{|\hbar \bar \omega'(p)|} \theta(\hbar \omega_j(p)<\epsilon_F),\label{Main}
\end{align}
where 
\begin{align}
|\hbar \bar \omega'(p)|\equiv \frac{\hbar^2}{ma^2}|\sin(p)|.
\end{align}

We now discuss the continuum limit of the dispersion relation. The function $\tilde{\alpha}(\omega)$ can be written in the continuum limit ($N\to\infty,$ $\Delta \sim  1/N$) as follows:\begin{align}&\tilde{\alpha}(\omega)=\hbar\alpha^2 \left(\dashint_{\omega_<}^{\omega_>}\frac{ H^2(\Omega)}{\omega-\Omega  }d\Omega+\delta(\omega) \right),\label{Continuumtildealpha}\end{align}where the integral is the Cauchy principal value integral and $H(\Omega)$\ is an interpolating function of $H_n$:
\begin{align}
&H( G_n)\equiv  H_n,\end{align}
while $\delta(\omega)$  (not to be confused with Kronecker's delta) {is a function that appears when taking the continuum limit, $N\to\infty,$ $\Delta \sim  1/N,$ of Eq. (\ref{tildealpha}).} The function  depends on the microscopic location of $\omega$ within the reservoir frequencies, $G_n$, and is given by:  
\begin{align}
&\delta (\omega)=\left\{ \begin{array}{lr} \pi H^2(\omega) \cot \frac{(\omega-\omega_<)\pi}{\Delta } & \omega\in[\omega_<,\omega_>], \quad |\omega-\omega_{<,>}|\gg\Delta  \\
0 &\omega\notin[\omega_<,\omega_>], \quad |\omega-\omega_{<,>}|\gg\Delta
 \end{array}\right. .
\end{align}

The $N+1$ solutions of the dispersion relation, Eq. (\ref{DispersionRelation}) in the small $\alpha$ limit, are composed of $N$\ perturbed reservoir energy levels and $1$ perturbed non-reservoir fermionic level. If, for example, the reservoir is above the unperturbed Fermi sea, $\epsilon_F<\hbar \omega_<<\hbar \omega_>,$ the perturbed reservoir energies will typically also lie above the Fermi sea. In this case only the perturbed energy of the non-reservoir fermion will lie under the Fermi sea. Since this state is below the reservoir   $\delta(\omega)=0 $ must be set in Eq. (\ref{Continuumtildealpha}).  One can then have that only the perturbed non-reservoir level lies below the Fermi energy $\epsilon_F$. This leads to the following form for the fermionic occupation: 
\begin{align}
\<c^\dagger (p)c(p)\>=\frac{f(e^{\imath p})+1}{2} , \quad f(e^{\imath p})=-1\mbox{ for }|p|>p_F\label{ffromc}
\end{align}   
where $f(e^{\imath p})$ is an arbitrary function of $p$ for $|p|<p_F$. The Fermi momentum $p_F$ is defined  as the momentum for which the dispersion relation, Eq. (\ref{DispersionRelation}), is solved with $\omega=\omega_F\equiv\epsilon_F/\hbar$:
\begin{align}
\hbar\omega_F =\varepsilon(p_F)+\tilde \alpha(\omega_F),
\end{align}

More generally,  it is easy then to find different models of the reservoir, defined by different values of $H_n,$  $\omega_<$, $\omega_>$ and $\Delta$ such that Eq. (\ref{ffromc}) holds. The Fermi momentum,
$p_F,$ is defined such as that the solution with $\delta(\omega)=0$ of (\ref{DispersionRelation})  exists with $p=p_F$ and $\hbar\omega=\epsilon_F$.  The combination $\frac{f(e^{\imath p})+1}{2}$ is chosen for later convenience. 

Other possibilities of course exist generalizing Eq. (\ref{ffromc}), where the distribution inside and outside the Fermi sea is arbitrary, but we shall not try to describe the reservoirs leading to such distribution here for the sake of brevity. The problem of finding such reservoirs was treated in other works \cite{Chin:Rivas:Huelga:Plenio:Exact:Mapping:Reservoirs,Nuesseler:Dhand:Ish:Huelga:Simulation:Open:Systems:Fermions}. {We note that although for our choice of reservoir above the Fermi sea, the specific functional form of the coupling function $H(\omega)$ plays little role, in other cases this function may introduce singularities into the ferminic occupation number and thus play a more crucial role in determining the behavior of entanglement in the system.}  What is rather general is that there is a jump discontinuity in the value of  $\<c^\dagger (p)c(p)\>.$ This is the general case if $|\hbar \omega_j'(p)|\neq 0.$ If $|\hbar \omega_j'(p)|= 0,$ one the other hand,  a power-law behavior around the Fermi momenutm rather than a jump discontinuity occurs. We shall treat here the problem where the behavior around the Fermi point is of a jump discontinuity.

To simplify the analysis later on we shall assume that the two Fermi points, $\pm p_F$ are symmetric, namely:
\begin{align}
f(e^{\imath (p_F+\delta p)})=f(e^{\imath (-p_F-\delta p)}),  
\end{align}
which can be more succinctly written as:
\begin{align}
f(z)=f(z^{-1}). \label{FPsymmetry}
\end{align}

\section{Entanglement Entropy and Negativity}
We consider a translationally invariant Gaussian fermionic systems with density matrix\begin{align}\rho=\frac{1}{Z} e^{\sum_{ij}\Omega_{ij} c^\dagger_i c_j}, \,\tr\rho=1\end{align}on the infinite line. The covariance matrix $f_{i-j}$ is defined by:
\begin{align}
f_{i-j}=\tr \left((2\rho-\mathds{1}) c^\dagger_ic_j\right),
\end{align}
which leads naturally to define the symbol $f(z)$, namely the generating function for $f_{i-j}$:
\begin{align}
f(z)=\sum_i f_i z^i.
\end{align}
Note that this definition leads to the following expression for the average occupation fermionic occupation number, $\<c^\dagger(p)c(p)\>,$ as a function of momentum, $p$:\begin{align}
\<c^\dagger(p) c(p)\>\equiv\tr \left(\rho c^\dagger(p)c(p)\right)=\frac{f(e^{\imath p})+1}{2},
\end{align}
conforming with Eq. (\ref{ffromc}).  We must deal then with the problem of $f(p)$ having a jump discontinuity, due to the presence of a Fermi surface, where namely a momentum $p=p_F$ at which the function $f(e^{\imath p})$ jumps (for higher dimensions the dimensionality of the Fermi surface also plays an important role, see Ref. \cite{Cramer:Eisert:Plenio:Statistics:Dependence}).

The density matrix and the covariance matrix are of course related to each other. {The density matrix  may be diagonalized in a multi-particle basis based on a single-particle basis that diagonlizes the covariance matrix.} We get an eigenvalue of the density matrix for every choice of whether a fermion occupies one of the fermionic eigenstates, that diagonalize  the covariance matrix \cite{Peschel:Correlation}. Thus if there are $N$\ eigenvalues of the covariance matrix one finds $2^N$ eigenvalues of the density matrix. Specifically, if $\lambda_i$ are the $N$ eigenvalues of the covariance matrix, then for any choice of $N$ values  for $\nu_i$, where $\nu_i\in\{\frac{1+\lambda_i}{2},\frac{1-\lambda_i}{2}\}$ (corresponding to the choice of occupied and unoccupied fermionic states, respectively),   there will correspond an eigenvalue of the density matrix, $\rho,$  of the form:
\begin{align}
\prod_{i} \nu_i.
\end{align}
Despite of the fact that the eigenvalues of the density matrix are the \emph{ products }of the $\nu_i$, we  shall refer, in a somewhat loose manner, to the \emph{individual} values of $\nu_i,$ as the `eigenvalues of the density matrix' and denote them, as we have just done, by $\nu_i$. The spectrum of $\nu_i$ is known as the entanglement spectrum, and is related to the spectrum of $\lambda_i$ as follows:
\begin{align}
\pm\lambda_i=2\nu_i-1.\label{eigenrelation}
\end{align} 
 The $\pm$ sign allows one to obtain both eigenvalues of the density matrix,  $\nu_i$ and $1-\nu_i,$ from the same value of $\lambda_i$. If the spectrum of the covariance matrix, $\lambda_i$, is known, or equivalently,  the entanglement spectrum, $\nu_i, $is known, then  the entropy of the system may be computed through
\begin{align}
S=\sum_i \nu_i\log(\nu_i)+(1-\nu_i)\log(1-\nu_i)=\sum_i \frac{1+\lambda_i}{2}\log \frac{1+\lambda_i}{2}+\frac{1-\lambda_i}{2}\log \frac{1-\lambda_i}{2}.
\end{align}

Translational invariance makes the Fourier transform $f(z)$ of the correlation function $f_{i-j}$\ a  very convenient object to work with. Indeed,  for any linear combination of fermionic operators $c^{(a)}=\sum a_i c_i$  one forms the function $g^{(a)}(z)=\sum a_i z^{-i}$ and one easily obtains:
\begin{align}
h^{(a)}(z)\equiv\sum_i z^{i} \tr \left((\mathds{1}-2\rho )c^\dagger_i c^{(a)}\right)=f(z)g^{(a)}(z).\label{fmultipiclative}
\end{align}  
$h^{(a)}$ encodes as a Fourier transform the correlation function of $c^{(a)}$ with any fermionic operator $c^\dagger_i$. We see that $f$ acts in a simple multiplicative fashion
 on $g^{(a)}$ to yield $h^{(a)}$, which is of great convenience. 

\begin{figure}[h!!!]
\begin{center}
\includegraphics[width=10cm]{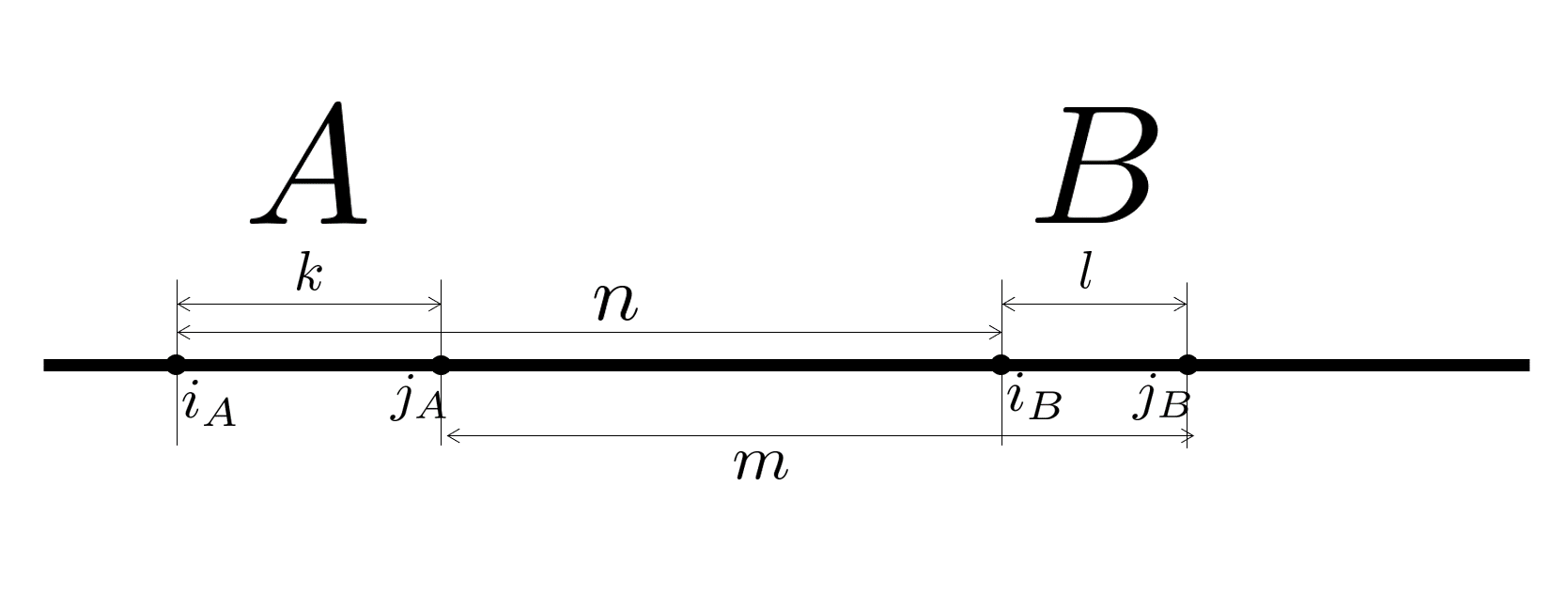}
\caption{
Two subsystems $A$ and $B$\ are identified on the 1 dimensional  real line (heavy line). The size of $A$ is $k+1$\ and its endpoints are $i_A$ and $j_A$ while the size of $B$\ is $l+1$ and its endpoints are $i_B$ and $j_B$. The distance $n$ is defined as $n=i_B-i_A$ while $m$ is defined as $j_B-j_A$.  \label{System}} \end{center}
\end{figure} 
Take now two intervals $A$ and $B$ on the infinite line as shown in Fig. \ref{System},  with $A$\ to the left of $B$. Let $k+1$ be the size of $A$ and $l+1$ be the size of $B$ and the distance between the leftmost point of $A$\ and the leftmost point of $B$ being $n$. Let $i_A$ bet the leftmost point in $A$ and $i_B$ be the leftmost point in $B$. Define also:

\begin{align} k_j=j-\left\{\begin{array}{lr} i_A  & j\in A\\ i_B & k\in B\end{array}\right.
.\end{align}

We now write an analogue to Eq. (\ref{fmultipiclative}). To this aim first take any fermionic operator $c^{(ab)}=c^{(a)}+c^{(b)}$ where $c^{(a)}$ is a linear combination of fermionic operators in $A$ and $c^{(b)}$ is a linear combination of fermionic operators in $B:$
\begin{align}
c^{(a)}=\sum_{j\in A}a^{(a)}_j c_j,\quad c^{(b)}=\sum_{j\in B}a^{(b)}_j c_j.
\end{align}
Now define $g_m^{(ab)}$ where $m=1,2$ as:
\begin{align}
g_1^{(ab)}=\sum_{j\in A} a^{(a)}_j z^{-k_j},\\
g_2^{(ab)}=\sum_{j\in B} a^{(b)}_j z^{-k_j}.
\end{align}
We wish to find the correlator of $c^{(ab)}$ with any fermionic operator in $A$\ or in $B$. To this aim we define $h_{1}^{(ab)}, $ $h_{2}^{(ab)}$ that encode these correlators as a Fourier transform:
\begin{align}
&h_1^{(ab)}(z)\equiv\sum_{i} z^{i-i_A} \tr \left((\mathds{1}-2\rho) c^\dagger_ic^{(a)}\right)
,\\ &h_2^{(ab)}(z)\equiv\sum_{i} z^{i-i_B} \tr \left((\mathds{1}-2\rho) c^\dagger_i c^{(b)}\right).
\end{align}
Then one has the following relations:
\begin{align}
&h^{(ab)}_1(z)=f(z)g^{(ab)}_1+f(z)z^n g_2^{(ab)}\\   
&h^{(ab)}_2(z)=f(z)z^{-n}g^{(ab)}_1+f(z) g_2^{(ab)}. 
\end{align}
These can be written in  matrix form:
\begin{align}
\bm{h}^{(ab)}=\bm f {\bm g}^{(ab)}, \quad \bm f(z)=\begin{pmatrix}f(z) & f(z)z^n \\
f(z)z^{-n} & f(z)\\
\end{pmatrix}.
\end{align}
The first equation here is the generalization of Eq. (\ref{fmultipiclative}) that we seek, where $f$ is replaced by the matrix $\bm f$. Thus, $\bm f$ will be the\ object describing the correlations within and between the two regions $A$ and $B$. 

One may look for the reduced density matrix namely a Gaussian matrix $\rho^{AB}$ such that for any $i,j\in A \cup B$ 
\begin{align}
\tr \rho^{AB}c_i^\dagger c_j=\tr \rho c^\dagger_ic_j, \quad \rho^{AB} =\frac{1}{Z^{AB}} e^{\sum_{i,j\in A \cup B}\Omega^{AB}_{ij}c^\dagger_ic_j}, \quad \tr\rho^{AB}=1.
\end{align}
Since the right hand side of the first equation is determined by $f_{i-j}$ restricted on $A\cup B$ and the left hand side by $\Omega^{AB}$, the problem reduces to finding $\Omega^{AB}$ given $f_{i-j}$. In fact, both matrices are diagonalized in the same basis, and there is a simple relation between their eigenvalues, Eq. (\ref{eigenrelation}) still holds. The matrix $f_{i-j}$  restricted to $A\cup B$\ is in turn encoded in $\bm f$. An eigenfunction $ \bm\psi^{(-\lambda)}$ with eigenvalue $-\lambda^{(kl)}_i,$ where $i=1,2,\dots,k+l$ solves the following equation:
\begin{align}
\mathcal{P}_{kl}(\bm f+\lambda_i^{(kl)}\mathds{1}) \bm\psi^{(-\lambda^{(kl)}_i)}=0,\label{evEqnotilde}
\end{align} 
where $\mathcal{P}_{kl}$ is the projection operator onto $A\cup B$:
\begin{align}
\mathcal{P}_{kl}\sum_i \bm\left(\begin{array}{c} a_{1i}z^i\\a_{2i}z^i \end{array} \right) =\left(\begin{array}{c}\sum_{0\leq i < |A|}a_{1i}z^i \\\sum_{0\leq i < |B|}a_{2i}z^i  \end{array}\right)\end{align}  
and $\bm \psi^{(-\lambda)}$ is in $A\cup B$, namely $\mathcal{P}_{kl}\bm \psi^{(-\lambda)}=\bm \psi^{(-\lambda)}$. This requires $\psi_1$ and $\psi_2$ to be polynomials of degree $|A|-1$ and $|B|-1$ respectively. So, in effect, one searches for orthogonal polynomials with respect to the measure $\bm f$. Once the eigenvalues are found, the eigenvalues $\nu_i^{(kl)}$ of $\Omega^{AB}$ follow from Eq. (\ref{eigenrelation}), repeated here for completion:
\begin{align}
\pm\lambda^{(kl)}_i=2\nu^{(kl)}_i-1.
\end{align} 
Then for example, the reduced or entanglement entropy of the region $AB$ can then be computed as follows:
\begin{align}
S^{AB}=\sum_i \frac{1+\lambda^{(kl)}_i}{2}\log \frac{1+\lambda^{(kl)}_i}{2}+\frac{1-\lambda^{(kl)}_i}{2}\log \frac{1-\lambda^{(kl)}_i}{2}.
\end{align}

As mentioned in the introduction one may wish to compute  the negativity spectrum. In this case one replaces the covariance matrix $f_{k_i,k_j}$ by  a deformed covariance matrix $\tilde{f}_{k_i,k_j}$\cite{shapourian:Hasan:Shiozako:ken:Ryu:Shinsei:Negativity,Shapourian:Ruggiero:Ryu:Twisted:Untwiste:Negativity}:

\begin{align}
\tilde f_{k_i,k_j}=\tr \left((\mathds{1}-2\rho )c^\dagger_i c_j\right)\times\left\{\begin{array}{lr}
1 & i\in A, j\in A\\
-1 & i\in B, j\in B\\
i & i\in A, j\in B\\
i & i\in B, j\in A\\ 
\end{array}\right.\label{deformingf} .
\end{align}
The problem is to find $\tilde{\rho}^{AB}$ such that 
\begin{align}
\tilde f_{k_i,k_j}=\tr \left((\mathds{1}-2\tilde \rho^{AB}) c^\dagger_i c_j\right), \quad \tilde{\rho}^{AB}=\frac{1}{\tilde{Z}^{AB}} e^{\sum_{i,j\in A\cup B}\tilde \Omega^{AB}_{i,j}c^\dagger_ic_j }, \quad \tr \tilde \rho^{AB}=1. 
\end{align}
 An eigenfunction $ \bm\psi^{(-\tilde \lambda^{(kl)}_i)}$ with eigenvalue $-\tilde \lambda_i^{(kl)}$  solves the following equation:
\begin{align}
\mathcal{P}_{kl}(\tilde{\bm{f}}+\tilde \lambda_i^{(kl)}\mathds{1}) \bm\psi^{(-\tilde\lambda^{(kl)}_i)}=0,\label{evEq}
\end{align} 
where, in accordance with Eq. (\ref{deformingf}), one has:
\begin{align}
\tilde {\bm{f}}(z)=\begin{pmatrix}f(z) & \imath f(z)z^n \\
\imath f(z)z^{-n} & - f(z)\\
\end{pmatrix} .
\end{align}
Once the roots of Eq. (\ref{evEq})\ are found one may find the  negativity spectrum through: 

\begin{align}
\pm\tilde \lambda^{(kl)}_i=2\tilde\nu^{(kl)}_i-1.\label{nulambdanega}
\end{align} 
{The} logarithmic negativity, written using the symbol $\mathcal E $ may be used as a potential that quantifies the entanglement of the system\cite{shapourian:monotone}.  It is computed  by considering only negative values of $\tilde{\rho}^{AB}$ and computing the sum of the logarithm of only those eigenvalues:
\begin{align}
\mathcal{E}=\sum_{i} \log(\left|\tilde  \nu_i\right|+\left| 1-\tilde     \nu_i\right|)=\sum_{i} \log\left(\left| \frac{1-\tilde \lambda_i}{2}\right|+\left| \frac{1+\tilde \lambda_i}{2}\right|\right)=\sum_{\{i||\tilde\lambda_ i|>1\}}\log|\tilde{\lambda}_i|.\label{negativity}
\end{align}
Indeed, if both $\tilde{\nu}_i$ and $1-\tilde \nu_i$ are positive, the sum, $\left|\tilde  \nu_i\right|+\left| 1-\tilde     \nu_i\right|,$ just gives $1$ and the contribution (after taking the logarithm) of the term associated with his expression in the sum after the first equality sign cancels leading to the fact that the logarithmic negativity only receives contributions that are associated with negative eigenvalues of  $\tilde{\rho}^{AB}$ .

\section{Orthogonal polynomials for two intervals} 

To accommodate both the computation of the entanglement entropy and logarithmic negativity we define a variable $\tau$ to be set as  $\tau=1$ when computing the entanglement entropy set as $\tau=\imath$ when  computing   logarithmic negativity. Namely,
\begin{align}
f_{ij}^{(\tau)}=\begin{cases}f_{ij} & \tau=1 \\
\tilde f_{ij} & \tau=\imath \\
\end{cases}.\label{tildefref}
\end{align}

To find the spectrum of $\lambda$ in Eqs. (\ref{evEq}, \ref{evEqnotilde}) one may resort to the usual method of taking the determinant $\det[\mathcal{P}_{kl}({\bm{f}}^{(\tau)}+ \lambda^{(\tau)}\mathds{1})]$ and finding those ${\lambda}^{(\tau)}$ . This leads one to define the matrix 
\begin{align}
{\bm f}^{(\tau)}(z)=\left(\begin{array}{cc}
 \lambda+f(z) & \tau  z^{n}f(z)\\
\tau z^{-n}f (z) & \lambda+\tau^2 f(z)
 \end{array} \right).
\end{align}
\subsection{Orthogonal polynomials} We shall need orthogonal polynomials of four types enumerated by the subscript $\sigma$ and $\omega$ where
$\omega\in\{+,-\}$ and $\sigma\in\{1,2\}.$  We write the following symbol ${\bm \psi}^{kln}_{\sigma \omega}$ for these two-dimensional vectors of polynomials where the superscript denotes the fixed values related to the position and size of the two regions $A$\ and $B.  $ Each vector has two components thus we have:
\begin{align}
{\bm \psi}^{kln}_{\sigma \omega}=\left(\begin{array}{c}
\psi^{kln}_{1 \sigma \omega}\\
 \psi^{kln}_{2\sigma\omega}\end{array}\right).
\end{align}
The elements of the vector satisfy:
\begin{align}
\label{Pchidef} \psi^{kln}_{\sigma \omega}(z) \mbox{ }\begin{array}{lr}
\mbox{ is monomial of degree }m_\sigma  & \mbox{if } \omega=+1 \\ \\ 
\mbox{satisfies } \psi^{kln}_{\sigma \sigma \omega}(0)=\chi^{-1}_{kln\sigma -}& \mbox{if } \omega=-1\\
\end{array},
\end{align}
where
 \begin{align}
&  m_1=k,\,m_2=l. \label{ms}
\end{align}
The functions $\psi^{kln}_{\alpha\sigma\omega}(z)$ for $\alpha\neq\sigma$ are  polynomials of degree $m_\alpha$. 

The following property of $\bm \psi$ makes it into a vector orthogonal polynomial:\begin{align}
&\bm{e}_{\sigma'}\int  z^{-j }{\bm f}^{(\tau)}(z;\lambda) {\bm \psi}^{kln}_{\sigma \omega}(z) \frac{d\theta}{2\pi} =\chi^{\frac{\omega+1}{2} }_{kln\sigma\omega}\delta_{\sigma,\sigma'}\delta_{j,(\omega+1) m_\sigma}\label{chidef}
\end{align}
 here $\frac{\omega+1}{2}$ is an exponent not an index, and\ $\bm e_\sigma$ is the unit vector in direction $\sigma,$ namely the $\sigma'$ element of $\bm e_\sigma$, denoted by $e_{\sigma\sigma'}$ is given by:
\begin{align}
e_{\sigma \sigma'}=\delta_{\sigma\sigma'}.
\end{align} 
 
\subsection{Computation of Characteristic Polynomial by Means of the Orthogonal Polynomials}The characteristic polynomial, namely the determinant of the operator $\mathcal{P}_{kl}(\lambda+ {\bm f}_n^{(\tau)}),$ is to be denoted as follows:
\begin{align}
D_{kln}=\det\mathcal{P}_{kl}(\lambda+ {\bm f}^{(\tau)})\label{lambdadefdef}=\prod_{i=1}^{k+l} \left(\lambda + \lambda_i^{(\tau kl)}\right),
\end{align}
where we used the notation  $\lambda_i^{(\tau kl)}$ standing for:
\begin{align}
\lambda_i^{(\tau kl)}=\begin{cases}\lambda_i^{( kl)} & \tau=1 \\
\tilde\lambda_i^{( kl)} & \tau=\imath  \\
\end{cases},\label{lambdataukl}
\end{align}
where $\lambda_i^{(kl)}$ and $\tilde \lambda_i^{(kl)}$ appeared before in Eqs. (\ref{evEqnotilde}, \ref{evEq}). We now give the following equations that relate the determinants that compute the characteristic polynomial  (once the  $\chi_{kln\sigma\omega}$'s defined in Eq. (\ref{Pchidef}, \ref{chidef}) are given) that is obtained by means of the orthogonal polynomials:
\begin{align}
&  \quad\frac{D_{k,l,n}}{D_{k,l-1,n}}=\chi_{kln2+,\quad } \frac{D_{k-1,l,n}}{D_{k,l,n}}=\chi_{kln1+.}\label{DetsAndChis}
\end{align}

We show these relations   by identifying $\chi$  as diagonal elements  in an upper-lower triangular decomposition of the two-interval covariance matrix $f^{(\tau)}_{ij} $ of Eq. (\ref{tildefref}). Actually, the matrix to be decomposed is  $\mathcal{F}_{k+l+2}^{(\tau)}$ defined by $\mathcal{F}_{k+l}^{(\tau)}=\mathds{1}\lambda+f^{(\tau)},$  where we have explicitly denoted the size of the matrix in the subscript of  $\mathcal{F}_{k+l+2}^{(\tau)}$ .  We  shall write  $N=k+l+2$. The upper-lower triangular decomposition of  $\mathcal{F}_{k+l+2}^{(\tau)}$ takes the following form
\begin{align}
&\mathcal{F}_N^{(\tau)}  B_N=C_N,\mbox{   with}\\ 
&\mathcal{F}_N=\begin{pmatrix}\lambda+f^{(\tau)}_{11} & f^{(\tau)}_{12} & . & . \\
f^{(\tau)}_{21} & \lambda+f^{(\tau)}_{22} & . & . \\
. & . & . & . \\
. & . & . & \lambda+f^{(\tau)}_{NN} \\
\end{pmatrix},\mbox{   }B_N=\begin{pmatrix} 1 & . & . & . \\
0 & 1 & . & . \\
. & . & . & . \\
0 & 0 & 0 & 1 \\
\end{pmatrix}\mbox{ and}\\&C_N=\begin{pmatrix} \eta_{N,1} & 0 & 0 & 0 \\
. & \eta_{N,2} & 0 & 0 \\
. & . & . & . \\
. & . & . & \eta_{N,N} \\
\end{pmatrix}\nonumber.
\end{align}
The determinant of $\mathcal{F}^{(\tau)}_N$ is given by:
\begin{align}
\det \mathcal{F}_N^{(\tau)} =\prod_{i=1}^{N}\eta_{N,i}.
\end{align}  

We denote the $i$-th column of $B_N$  by $B_N^{(i)}$ , where $1\leq i\leq N$. It obeys the equations:
\begin{align}
&B^{(i)}_{N,i}=1\label{BNmonomial}\\
&B^{(i)}_{N,j}=0\mbox{ for } j>i\label{BNPolynomial}\\
&(\mathcal{F}_N^{(\tau)}  B_N^{(i)})_j=0\mbox{ for }j<i,\label{BNResult}
\end{align} 
where $B^{(i)}_{N,j}$ is the $j$-th element of $B_N^{(i)}$. The determinant is then given by 

\begin{align}
\det \mathcal{F}_N^{(\tau)} =\prod_{i=1}^{N}C^{(i)}_{N,i}\equiv \prod_{i=1}^{N}\eta^{}_{N,i}. \end{align}

We now show that $B^{(i)}_{N}$ is independent of $N$, namely $B^{(i)}_{N}=B^{(i)}_{N'}$, for $i<N,N'$. Indeed, Eqs. (\ref{BNmonomial}-\ref{BNResult})\ are independent of $N$. To see this, let $M>N,N'$, then, due to   Eq. (\ref{BNPolynomial}) , we may write the same equation for both  for both $B^{(i)}_N$\ and $B^{(i)}_{N'}$:
\begin{align}
\sum_{p=1}^i\mathcal{F}^{(\tau)}_{M, jp}B^{(i)}_{N,p}=0,\quad  \sum_{p=1}^i\mathcal{F}^{(\tau)}_{M, jp}B^{(i)}_{N',p}=0,\quad\mbox{ for } j<i.
\end{align} 
Since $B_N^{(i)}$ and $B_{N'}^{(i)}$ obey the same equations, they are equal.  As a result, $\eta_{N,i}=C^{(i)}_{N,i}=\sum_{1}^i\mathcal{F}^{(\tau)}_{M ,ip}B^{(i)}_{N,p}$ does not depend on $N$ as well and we may just write $\eta_i\equiv \eta_{N,i}$ for any $N\geq i$. This fact allows us to write:
\begin{align}
\frac{\det \mathcal{F}_N^{(\tau)}}{\det \mathcal{F}^{(\tau)}_{N-1}}= \frac{\prod_{i=1}^N \eta_{N,i}}{\prod_{i=1}^{N-1} \eta_{N-1,i}}=\frac{\prod_{i=1}^N \eta_{i}}{\prod_{i=1}^{N-1} \eta_{i}}=\eta_N.
\end{align}
To obtain the first equation in  (\ref{DetsAndChis}) we have to show that $C^{(N)}_{N,N}=\chi_{kln2+.}$ To do so note that one may identify
\begin{align}
 {\bm \psi}^{kln}_{2 +}(z)=\begin{pmatrix}\sum_{i=0}^{k} z^iB_{N,i}^{(N)}\\
\sum_{i=0}^{l} z^iB^{(N)}_{N,i+k+1}\\
\end{pmatrix},
\end{align} 
This identification being made  by realizing Eqs.(\ref{BNPolynomial}-\ref{BNResult})
being the Fourier transform of (\ref{chidef}) and Eq. (\ref{BNmonomial}) being the first line of Eq. (\ref{Pchidef}).
Under this identification, $C^{(N)}_{N,N}$ becomes equivalent to $\chi_{kln2+}.$

It is possible to transpose the intervals $A$\ and $B$, whereupon one obtains the second equation in   (\ref{DetsAndChis}) analogously to how we just obtained the first one. 
\section{Riemann-Hilbert Problem for Orthogonal Polynomials}

{Having defined the orthogonal polynomials, one may find their asymptotic behavior by associating with them a Riemann-Hilbert problem. In particular, one writes a matrix $T(z),$ the elements of which feature the orthoghonal polynomials (Eq. (\ref{Tdef}) below).  The analytic properties of these matrix as a function of $z$ are, on the one hand, encode the orthogonality properties of the polynomials, and, on the other hand, naturally define a matrix Riemann-Hilbert problem. Solving approximately the Riemann-Hilbert problem (something that we are able to do here only in some limits) then allows to find the asymptotes of the orthogonal polynomials. These latter encodes in turn the entanglement spectrum of the fermionic system. }

To find  the behavior of  the orthogonal polynomials, ${\bm \psi},$ at large $k,$ $l$ and $n$, one defines first the matrix $T(z)$:
\begin{align}
&T(z)\label{Tdef}=\\&=\nonumber
\left(\begin{array}{cccc}
 \psi_{11+}^{k(l-1)n} & \psi_{21+}^{k(l-1)n} & \oint\frac{ {-\bm e}_2 {\bm f}^{(\tau)}(\xi;\lambda) {\bm \psi^{k(l-1)n}_{1+}}(\xi)}{(z-\xi)2\pi\imath\xi^l}  & \oint\frac{ {-\bm e}_1 {\bm f}^{(\tau)}(\xi;\lambda) {\bm \psi^{k(l-1)n}_{1+}}(\xi)}{(z-\xi)2\pi\imath\xi^k} \\
\psi_{12+}^{(k-1)ln} &  \psi_{22+}^{(k-1)ln} &   \oint\frac{ {-\bm e}_2 {\bm f}^{(\tau)}(\xi;\lambda) {\bm \psi^{(k-1)ln}_{2+}}(\xi)}{(z-\xi)2\pi\imath\xi^l}  &  \oint\frac{ {-\bm e}_1 {\bm f}^{(\tau)}(\xi;\lambda) {\bm \psi^{(k-1)ln}_{2+}}(\xi)}{(z-\xi)2\pi\imath\xi^k} \\
-\psi_{12-}^{(k-1)(l-1)n} & -\psi_{22-}^{(k-1)(l-1)n} &  \oint\frac{ {\bm e}_2 {\bm f}^{(\tau)}(\xi;\lambda){\bm\psi}^{(k-1)(l-1)n}_{2-}(\xi)}{(z-\xi)2\pi\imath\xi^l}  &  \oint\frac{ {\bm e}_1 {\bm f}^{(\tau)}(\xi;\lambda){\bm\psi}^{(k-1)(l-1)n}_{2-}(\xi)}{(z-\xi)2\pi\imath\xi^k} 
 \\
-\psi_{11-}^{(k-1)(l-1)n} & -\psi_{21-}^{(k-1)(l-1)n} &  \oint\frac{ {\bm e}_2 {\bm f}^{(\tau)}(\xi;\lambda){\bm\psi}_{1-}^{(k-1)(l-1)n}(\xi)}{(z-\xi)2\pi\imath\xi^l}  &  \oint\frac{ {\bm e}_1 {\bm f}^{(\tau)}(\xi;\lambda){\bm\psi}^{(k-1)(l-1)n}_{1-}(\xi)}{(z-\xi)2\pi\imath\xi^k} 
\end{array} \right).
\end{align}
Eventually we are interested in finding $\chi_{kln\sigma\omega}$ which are  encoded in \begin{align}&T_{41}(0)=\chi^{-1}_{k-1,l-1,1-},\label{TsandChis34} \quad T_{32}(0)=\chi^{-1}_{k-1,l-1,2-},\\&T_{23}(0)=\chi^{}_{k-1,l,2+},\quad T_{14}(0)=\chi^{}_{k,l-1,1+,\label{TsandChis12}}\end{align} 
as can be easily verified by comparing the respective  elements of the matrix $T(z)$ given in Eq. (\ref{Tdef}) with the definition of $\chi$ given in Eqs. (\ref{Pchidef},\ref{chidef}).

The matrix $T(z)$  has the following behavior at infinity (spaces here and below denote zero elements):
\begin{align}
T(z)=(\mathds{1}+O(1/z))\left(\begin{array}{cccc}
z^k & & &\\
&z^l & &\\
& & z^{-l} &\\
& & & z^{-k}
\end{array}\right)
\end{align}
and the following Riemann-Hilbert Property
\begin{align}
T_+(z)=T_-(z)\left(\begin{array}{cccc}
1&0& \tau f(z)z^{-l-n} &  (\lambda+f(z)) z^{-k}\\
0&1& (\lambda+\tau^2 f(z))z^{-l}& \tau f(z)  z^{-k+n} \\
0&0& 1&0 \\
0&0& 0 &1
\end{array} \right)
\end{align}
where $T_+(z)$ is the limiting value of  $T(z)$ on the unit circle approaching  from the interior of the circle, and $T_-(z)$ is the limiting value of  $T(z)$ on the unit circle approaching from the exterior. 
This matrix is further modified by applying the following transformation:
\begin{align}
&Y(z)\left(\begin{array}{cccc}
z^{k}&0&0 & 0\\
0&z^{l}& 0& 0 \\
0&0& z^{-l}&0 \\
0&0& 0 &z^{-k}
\end{array} \right)=T(z), \mbox{ for }|z|>1\\& Y(z)=T(z), \mbox{ for }|z|<1.
\end{align}
With this modification , the Riemann-Hilbert property of $Y$reads
\begin{align}
&Y_+(z)=Y_-(z) V\\&V=\left(\begin{array}{cccc}
z^k&0& \imath  f(z)z^{-m} &  \lambda+f(z) \\
0&z^l& \lambda- f(z)& \imath f(z)  z^{m} \\
0&0& z^{-l}&0 \\
0&0& 0 &z^{-k}
\end{array} \right)
\end{align}
where $m=n+l-k,  $  and $V$ will be termed 'the jump matrix'. The matrix $Y$ has the following behavior at infinity: 
\begin{align}
Y(z)\overset{z\to\infty}= \mathds{1}+O(1/z).\end{align}
\section{Solution for the Outer region}

The jump matrix, $V$, may be decomposed as follows:
\begin{align}
&\left(\begin{array}{cccc}
z^k&0& \tau  fz^{-m} &  \lambda+f \\
0&z^l& \lambda+\tau^2 f& \tau f  z^{m} \\
0&0& z^{-l}&0 \\
0&0& 0 &z^{-k}
\end{array} \right)=V^{(1)}V^{(2)}V^{(3)},\label{Lensdecomposition}   
\end{align}
where
\begin{align}
&V^{(1)}=V^{V,VI}=V^{V,IV}=\label{VVVI}\begin{pmatrix}1 & 0 & -\tau fz^{l-m} & 0 \\
0 & 1 & 0 & 0 \\
0 & \frac{-z^{-l}}{ \lambda+\tau^2 f} & 1 & 0 \\
\frac{-z^{-k}}{\lambda+f} & 0 & \frac{\tau f z^{-k+l-m}}{\lambda+f} & 1 \\
\end{pmatrix}^{-1}, \\&V^{(2)}=V^{IV,III}=V^{VI,I}\label{VII}=\left(\begin{array}{cccc}
0&0& 0 & \lambda+f\\0&0& \lambda+\tau^2 f& 0 \\
0&\frac{-1}{\lambda+\tau^2 f}& 0&0      \\
\frac{-1}{\lambda+f}&0& 0 & 0 
\end{array} \right),\\&V^{(3)}=V^{I,II}\label{VIII}=V^{III,II}=\begin{pmatrix}1 & 0 & 0 & 0 \\
0 & 1 & 0 &\tau f z^{m-l} \\
0 & \frac{z^l}{\lambda+\tau^2 f} & 1 &\frac{\tau f z^m}{\lambda+\tau^2 f} \\
\frac{z^k}{\lambda+ f} & 0 & 0 & 1 \\
\end{pmatrix}.
\end{align}
The meaning of the double roman superscripts will be explained in the following. 

\begin{figure}[h!!!]
\begin{center}
\includegraphics[width=10cm]{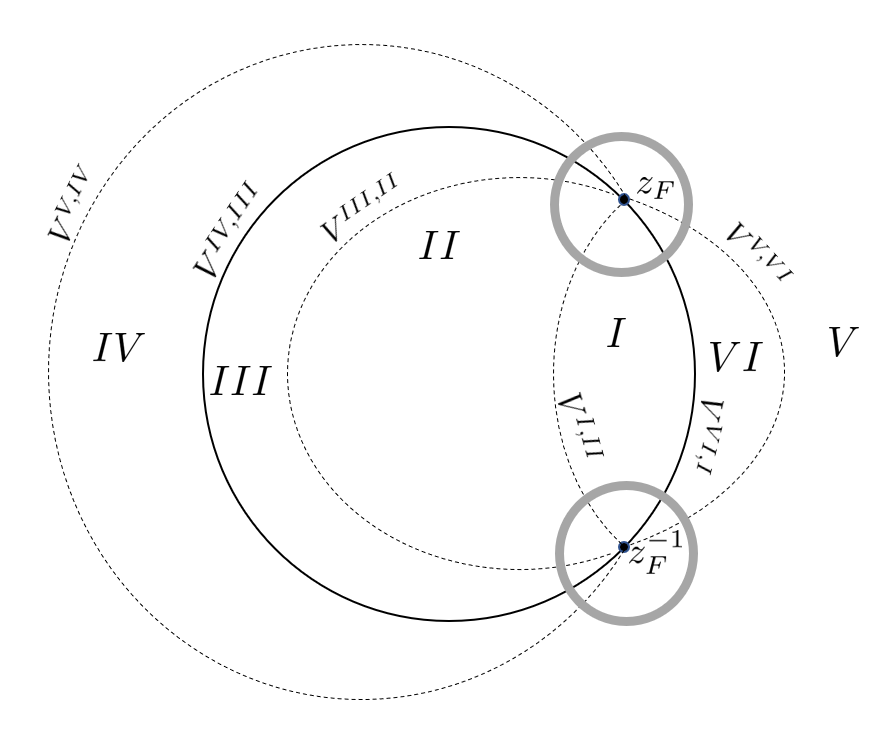}
\caption{
The decomposition of the complex plane into six regions. {The heavy line denotes the unit circle over which  the original jump matrix is defined. The dotted lines denote the lines over which new jump matrices are defined after decomposing the original jump matrix. Outside the gray circles the jump matrices on the dotted lines tend to the identity and thus are irrelevant, while the remaining jump matrices on the unit circle become trivial to solve.  Inside the gray circles one must take all the jump matrices into account. } \label{Decomposition}} \end{center}
\end{figure}

A decomposition of the form (\ref{SzegoComesfrom}) has been shown in Refs. \cite{Deift:Its:Krasovksy:Toeplitz:Hankel,Deift:Its:Krasovsky:Toeplitz,Deift:Zhou:Steepest:Descent} to allow for a large $N$ expansion (here $N$ is the order of $k$, $l$). We review this method presently. First, note that the function $f(z)$ has jump discontinuities at $z_F$ and $z_F^{-1}$, but otherwise is assumed to be smooth and analytic. Let us fix two circles around the points $z_F$ and $z_F^{-1}$ denoted by the heavy gray circles in Fig. \ref{Decomposition}. Removing these two circles, the unit circle decomposes into two arcs, the first one going clockwise from the gray circle surrounding $z_F$ to $z_F^{-1}$ and the second going counterclockwise. One assumes that the function $f$ has an analytic continuation from the first arc to a neighbourhood of the first arc and from the second arc to the neighbourhood of the second arc. 

One now draws 
 four lines -- the dashed lines in Fig. \ref{Decomposition} -- that connect $z_F$ and $z_F^{-1}$, with the condition, that when they leave the gray circles, they always remain in the neighbourhood of the arcs in which $f(z)$ has an analytic continuation as just discussed. 
The dashed lines and the unit circle together divide the plane into six regions which are denoted by Roman numerals as in Fig. \ref{Decomposition} The matrix $V^{(1)}$ may then be analytically continued all the way to the two  dashed lines outside the unit circle, these analytical continuations are denoted by $V^{V,VI}$ and $V^{V,IV},$ respectively.  a similar procedure can be applied to $V^{(3)}$ and the dashed lines inside the unit circle produces $V^{III,II}$ and $V^{I,II}$, respectively, while $V^{(2)}$ requires no analytic continuation since one continues to use this matrix on the unit circle, but we still denote it by two symbols $V^{VI,V}$ and $V^{IV,III}$ depending on which arc of the unit circle it is being used.

After analytic continuation of the Riemann-Hilbert problem, which originally had  jump discontinuities only on the unit circle now has jump  discontinuity on the system of six arcs ($4$ dashed arcs of Fig. \ref{Decomposition}, and $2$ arcs of the unit circle).  The solution of the new Riemann-Hilbert problem in the different regions is denoted by $Y_{I-VI}$ with the subscript which gives the region with which the solution is valid. The solution of the original Riemann-Hilbert problem  (the Riemann-Hilbert problem with jump matrix $V$) is denoted by $Y$. The solution of the new Riemann-Hilbert problem is written in terms of the original one as follows:
\begin{align}
&Y_{II}=Y,\quad Y_V=Y  \quad Y_{VI}=Y V^{V,VI}, \quad Y_{IV}=Y V^{V,IV}, \quad Y_{VI}=Y V^{V,VI}\label{Ysrelations}\\
& Y_{III}=Y V^{II,III}=Y_{IV} V^{IV,III}, \quad Y_{I}=YV^{II,I}=Y_{IV} V^{VI,I}.\nonumber
\end{align}  
It is easy to check that these relation are consistent on account of $Y$ obeying the original Riemann-Hilbert problem. 

The utility  of the procedure just outlined for the large $N$\ expansion ($k\to\infty$, $l\to\infty$, $m-l\to\infty$) is that on the dashed lines outside the gray circles, the jump matrices (the  matrices $V^{V,VI},$  $V^{V,IV},$ $V^{VI,V}$ and $V^{IV,III}$ )   all converge to the  unit matrices to exponential accuracy, as can be ascertained by examining the expression for those matrices in Eqs. (\ref{VVVI}, \ref{VIII}). Taking into account the equations in (\ref{Ysrelations}) this means that $Y_{I}=Y_{III}=Y_{II}=Y$ and $Y_{V}=Y_{IV}=Y_{VI}=Y,$ and that is true to exponential accuracy in the large $N$ limit outside the vicinity of $z_F^{\pm1}$  (outside the gray circles of Fig. \ref{Decomposition}). In addition, one obtains the that the jump matrix is given by $V^{(2)}$ of Eq. (\ref{VII}), or in other words $V^{VI,I}$ and $V^{IV,III}$ on the two arcs connecting  $z_F$ and $z_F^{-1}$. One then first solves the Riemann-Hilbert porblem outside the vicinities of $z_F^{\pm1}$ by solving the Riemann-Hilbert problem associated with $V^{(2)}$ in that region. We shall denote the solution of this Riemann-Hilbert problem as $Y^{(\rm out)}$. After this has been performed, one looks for a solution inside the vicinities of $z_F^{\pm1}$.

We first fix some notations.  We assume a jump discontinuity of $f(z)$ at $z_F$ and at $z^{-1}_F$ (we assume Eq. (\ref{FPsymmetry}), namely $f(z)=f(z^{-1}))$:
\begin{align}
f(z_F e^{\pm\imath 0^+})=f(z^{-1}_F e^{\mp\imath 0^+})=f_{o/i.}
\end{align}
Namely $f_{o/i}$ are the limiting values of $f$ as one approaches $z_F$ from the left or right, respectively or vice versa for the point $z_F^{-1}$. 

We define also:
\begin{align}
&\label{rbeta}r=\sqrt{(\lambda+f_o)(\lambda+f_i)},&\tilde r=\sqrt{(\lambda+\tau^2 f_o)(\lambda+\tau^2  f_i)},\\
&\nonumber \beta=\frac{1}{2\pi\imath}\log\frac{\lambda+f_i}{\lambda+f_o},&\tilde\beta=\frac{1}{2\pi\imath}\log\frac{\lambda+\tau^2 f_i}{\lambda+\tau^2 f_o} 
\end{align}
such that one decompose the relevant jump matrix in the following way,(a form which will immediately lead to a solution of the relevant Riemann-Hilbert problem in the outer region):        
\begin{align}
&\left(\begin{array}{cccc}
0&0& 0 & \lambda+f\\0&0& \lambda+\tau^2 f& 0 \\
0&\frac{-1}{\lambda+\tau^2 f}& 0&0      \\
\frac{-1}{\lambda+f}&0& 0 & 0 
\end{array} \right)\equiv \mbox{\bf a-diag}(\lambda+f,\lambda+\tau^2 f,-(\lambda+\tau^2 f)^{-1},-(\lambda+f)^{-1})\label{antidaigdecomp}=\\
&=\nonumber\mbox{\bf diag}\left(
\frac{x^{ \beta}}{F_-(z)}, \frac{x^{ \tilde{'\beta}}}{F^{(\tau)}_-(z)} ,x^{ -\tilde{\beta}}F^{(\tau)}_-,     {x^{ -\beta}}F_-(z)\right)\times 
\\&\times\nonumber\mbox{\bf a-diag}\left(
 x^{ \beta}e^{-\imath \pi\beta}F_+(z), x^{ \tilde{\beta}} e^{-\imath \pi\tilde{\beta}}F^{(\tau)}_+(z),-\frac{x^{ -\tilde{\beta}}e^{\imath \pi\tilde{\beta}}}{F^{(\tau)}_+(z)},
-\frac{x^{ -\beta^{}}e^{\imath \pi\beta}}{F_+(z)} \right) 
\end{align}
where $\mbox{\bf{ a-diag}}$ and $\mbox{\bf{ diag}}$ denote:
\begin{align}
&\mbox{\bf diag}(a,b,c,d)=\begin{pmatrix} a &  &  &  \\
 & b &  &  \\
 &  & c &  \\
 &  &  & d \\
\end{pmatrix}, \\
&\mbox{\bf a-diag}(a,b,c,d)=\begin{pmatrix}
 \, &  &  &a  \\
 &  & b &  \\
 & c &  &  \\
d &  &  &  \\
\end{pmatrix}
\end{align}
and  
\begin{align}
x(z)=\frac{z-z_F}{z-z_F^{-1}}.
\end{align}
 The function $F^{(\sigma)}_-(z),$ for $\sigma\in\{1,\tau\}$ is defined as a function that is analytic in the exterior of the unit circle  having the asymptote $F^{(\sigma)}_-(z)\to 1 $ at $z\to\infty$, while $F^{(\sigma)}_+(z)$ is analytic in the interior of the unit circle with the relation:
\begin{align}
F_+^{(\sigma)} (z)F_-^{(\sigma)}(z)=\frac{\lambda+\sigma^2f}{e^{\imath \pi \beta^{(\sigma)}}\theta_{FS}(z)+e^{-\imath \pi \beta^{(\sigma)}} (1-\theta_{FS}(z))},
\end{align} 
where $\theta_{FS}$ is the indicative function of the Fermi sea (taking the value $1$ in the Fermi sea and 0 outside). Here $\beta^{(1)}\equiv \beta$ and  $\beta^{(\tau)}\equiv \tilde \beta.$ 

Of course, the functions $F^{(\sigma)}_\pm$ can be easily calculated using the usual Wiener-Hopf decomposition: 
\begin{align}
2\pi\imath\log F^{(\sigma)}_\pm(z)=\oint\log\frac{\lambda+\sigma^2f(\xi)}{e^{\imath \pi \beta^{(\sigma)}}\theta_{FS}(\xi)+e^{-\imath \pi \beta^{(\sigma)}} (1-\theta_{FS}(\xi))} \frac{d\xi}{z-\xi},
\end{align}
this equation holding true for $z$ inside or outside the unit circle, respective with the choice of sign $\pm$ in  subscript in $F^{(\sigma)}_\pm $, while for other $z$'s one must apply an analytic continuation. The following relation will be useful:
\begin{align}
\log[ x(z)^\beta e^{-\imath \pi\beta}F_+(z)]\overset{z\to0}\longrightarrow \oint\log\left(\lambda+f(\theta)\right)\frac {d\theta}{2\pi}.\label{SzegoComesfrom}
\end{align}

The decomposition of Eq. (\ref{antidaigdecomp}) allows us to immediately write as a  solution to the Riemann-Hilbert problem, albeit a solution which   ignores the inner region. The solution reads as follow:
\begin{align}
&Y^{\rm(out)}=\mbox{\bf a-diag}\left(
 x^{ \beta}e^{-\imath \pi\beta}F_+(z), x^{ \tilde{\beta}} e^{-\imath \pi\tilde{\beta}}F^{(\tau)}_+(z),-\frac{x^{ -\tilde{\beta}}e^{\imath \pi\tilde{\beta}}}{F^{(\tau)}_+(z)},
-\frac{x^{ -\beta^{}}e^{\imath \pi\beta^{}}}{F_+(z)} \right),\label{Youtin}\quad|z|<1\\
&Y^{\rm(out)}=\mbox{\bf diag}\left(
\frac{x^{ \beta}}{F_-(z)}, \frac{x^{ \tilde{\beta}}}{F^{(\tau)}_-(z)} ,x^{ -\tilde{\beta}}F^{(\tau)}_-(z),     {x^{ -\beta}}F_-(z)\right),\quad|z|>1,
\end{align} 
the superscript $\rm(out)$ denoting that this solution is valid outside the gray circles of Fig. \ref{Decomposition}, namely in the outer region.
One may easily use Eq. (\ref{antidaigdecomp}) to verify that this indeed is a solution.  
\section{Solution for the Inner and Middle Regions}
It remains to solve the Riemann-Hilbert problem in the vicinity of the points $z_F^{\pm1}$. We do so only in the limit where $m-l\gg k,l, $ and  supply a method to obtain the large $N$\ limit in the result for negativity or the entanglement spectra in a power series of $|l/m|, |k/m| $ to any power. We provide explicit results for the leading order, but higher orders may be obtained by the method we present here to any power. 

In order to obtain this power series, we draw an additional circle around each of the points $z_F^{-1}$ of a radius which is much larger than $1/|m-l|$ but much smaller than $1/k$. We shall call the region inside these circles as the 'inner region'. The region between this circle and the gray circles in Fig. \ 
\ref{Decomposition}, we shall call the 'middle region'. We shall now search for a solution within the inner and middle regions that we shall denote by $Y^{(\rm in)}$ and $Y^{\rm(mid)},$ respectively.

\subsection{Solution in the middle region}

First let us turn our attention to the middle region.  Observing that in this region we may drop terms proportional to $z^{l-m}$ and $z^{l-m-k}$ outside the unit circle and terms proportional to $z^{m-l}$ and $z^{m}$ inside the unit circle in Eqs. (\ref{VVVI}, \ref{VIII}), respectively, leads us to write the following jump matrices in this region:
\begin{align}
&V^{(1)}=\begin{pmatrix}
1 & 0 & 0 & 0\\
0 & 1 & 0 & 0 \\
0 &-\frac{z^{-l}}{\lambda+\tau^2f}& 1 & 0 \\
-\frac{z^{-k}}{\lambda+f} & 0 & 0 & 1
\end{pmatrix}, \quad V^{(3)}=\begin{pmatrix}
1 & 0 & 0 & 0\\
0 & 1 & 0 & 0 \\
0 &\frac{z^{l}}{\lambda+\tau^2f}& 1 & 0 \\
\frac{z^{k}}{\lambda+f} & 0 & 0 & 1
\end{pmatrix}
\end{align}
while $V^{(2)}$ remains that of Eq. (\ref{VII})
.
One observes that in this region, the jump matrix decomposes into two spaces one spanned by $\bm{e}^{(1)},$ and $\bm{e}^{(4)}$ and the other space spanned by $\bm{e}^{(2)},$ and $\bm{e}^{(3)}$, where $\bm{e}^{(i)}$ is the unit vector in the $i$-th direction ($\bm{e}^{(i)}_j=\delta_{ij}$).
 Thus instead of a $4\times4$ Riemann-Hilbert problem we have two $2\times2$ Riemann-Hilbert problems and we may borrow for this region then the $2\times2$ matrix problem solved in Ref. \cite{Deift:Its:Krasovksy:Toeplitz:Hankel}.

In order to solve the problem, we search for the elements of $Y_{II}^{\rm (mid)}$ as these will contain all the information needed to obtain $Y^{(\rm mid)}$ in any region. The function  $Y_{II}^{\rm (mid)}$ has the following monodromy around the origin:
\begin{align}
\begin{pmatrix}1 & 0 & 0 & f_i-f_o  \\
0 & 1 & \tau^2(f_i-f_o) & 0 \\
0 & 0 & 1 & 0 \\
0 & 0 & 0 & 1 \\
\end{pmatrix}=
\begin{pmatrix}1 & 0 & 0 & z^kr2\imath\sin(\pi\beta)  \\
0 & 1 & z^l2\tilde r2\imath\sin(\pi\tilde \beta) & 0 \\
0 & 0 & 1 & 0 \\
0 & 0 & 0 & 1 \\
\end{pmatrix},
\end{align} 
which can be derived by multiplying out all the jump matrices as one goes around the origin. This is the first requirement on  $Y_{II}^{\rm (mid)}(z).$ A second requirement is that as $k|\zeta|,l|\zeta|\gg 1$, the function $Y_{II}^{\rm (mid)}(z) $ must match asymptotically $Y_{II}^{\rm (out)}(z).$ As a third requirement one must demand that the   $Y^{\rm (mid)}(z)$ for all other regions (namely, regions $I$ through $VI$ excluding $II$) matches asymptotically $Y^{(\rm out)}$. 

Let us give a solution to these three conditions and ascertain them below.
Such a solution is given by:
\begin{align}
& Y^{\rm(mid)}_{II}(z)\label{YIImid}=\\&\nonumber\begin{pmatrix}\scriptscriptstyle\frac{d}{r k^\beta}e^{\zeta k}Q^0_\beta(k\zeta) & 0 & 0 & \scriptscriptstyle \frac{d}{k^\beta} P^0_\beta(k\zeta) \\
0 &\scriptscriptstyle \frac{\tilde d}{\tilde rk^{\tilde{\beta}}}e^{\zeta l} Q^0_{\tilde\beta}(l\zeta) & \scriptscriptstyle \frac{\tilde d}{k^{\tilde{\beta}}}P^0_{\tilde{\beta}}(l\zeta) \\
0 &\scriptscriptstyle  \frac{-e^{-\imath2\pi{\tilde{\beta}}}\Gamma(1-{\tilde{\beta}})}{\tilde{d}\Gamma({\tilde{\beta}})l^{-\tilde{\beta}}}e^{\zeta l}Q^1_{\tilde{\beta}}(l\zeta) &\scriptscriptstyle \frac{-\tilde{r} e^{-\imath2\pi{\tilde{\beta}}}\Gamma(1-{\tilde{\beta}})}{\tilde{d}\Gamma({\tilde{\beta}})l^{-\tilde{\beta}}} P^1_{\tilde\beta}(l\zeta) \\
\scriptscriptstyle\frac{-e^{-\imath2\pi\beta}\Gamma(1-\beta)}{d\Gamma(\beta)k^{-\beta}}e^{\zeta k}Q^1_\beta(k\zeta) & 0 & 0 &\scriptscriptstyle\frac{  -re^{-\imath2\pi\beta}\Gamma(1-\beta)}{d\Gamma(\beta)k^{-\beta}} P^1_\beta(k\zeta) \\
\end{pmatrix},
\end{align}
where\begin{align}\label{Qdef}
Q^{i}_{\alpha}(\zeta)&=-e^{\imath2 \pi \alpha}\psi^1_{i-\alpha}(e^{-\imath\pi}\zeta)+\frac{\Gamma(\alpha+1-i)}{\Gamma(-\alpha+i)}e^{\imath2 \pi\alpha }e^{-\zeta}\psi^1_{1-i+\alpha}(\zeta)
\\P^i_\alpha(\zeta)&=e^{\pi\imath\alpha}\psi^1_{i-\alpha}(e^{-\imath\pi}\zeta),\label{Pdef}
\end{align}
and $\psi^b_{a}(\zeta)$ usually denoted by $\psi(a,b,\zeta) $ is the confluent hypergeometric function (see, e.g., references \cite{abramowitz:stegun,Deift:Its:Krasovksy:Toeplitz:Hankel,Its:Krasovsky:Gaussian:With:Kump}) while
\begin{align}
\zeta=\log(z/z_F^{\pm1})
\end{align}
and the sign depends on whether we are dealing with the middle region around $z_F^{\pm1},$ respectively.  The first requirement on the solution mentioned above is satisfied since, the functions $P$\ and $Q$, when considered as elements of a  row vector  
\begin{align}
&\left(Q^{i}_{\alpha}(\zeta),P^{i}_{\alpha}(\zeta)\right)
\end{align}
have monodromy
\begin{align}
\begin{pmatrix}
1 & 2\imath \sin(\pi\alpha) \\
0 & 1 \\
\end{pmatrix}
\end{align}
as $\zeta\to e^{2\pi\imath}\zeta$. 
This may be derived by using the following property of the hypergeometric confluent function: 
\begin{align}
\label{psimonodromy}\psi_a^c(e^{-2\pi\imath}\zeta)=e^{2\pi\imath a}\psi^c_a(\zeta)-\frac{2\pi\imath}{\Gamma(a)\Gamma(a-c+1)}e^{\pi\imath a} e^{\zeta}\psi^c_{c-a}(e^{-\imath\pi}\zeta).
\end{align}
We have also used the following following definition of $d$ and $\tilde{d}$:
\begin{align}
d=F^{}_- (z_F^{\pm1})\left(1-z_F^{\mp2}\right)^{\mp\beta}\quad \tilde d=F_-^{(\tau)}(z_F^{\pm1})  \left(1-z_F^{\mp2}\right)^{\mp\tilde \beta},\label{ddef}
\end{align}
and the signs $\pm$ are to be taken respectively on whether the expansion is made around $z_F$ or $z_F^{-1}$. 

The second requirement on the solution, namely its asymptotics as $k|\zeta|,l|\zeta|\gg 1$ is also easily seen to be satisfied, as in this limit $Y_{II}^{(\rm mid)}$ has the following behavior:

\begin{align}
& \label{Ymidin}Y^{\rm(mid)}_{II}(\zeta)\overset{\zeta\to\infty}\to\\&\nonumber\begin{pmatrix}\scriptscriptstyle \frac{d\Gamma(1+\beta)e^{\imath2\pi\beta}\zeta^{-\beta-1}}{r\Gamma(-\beta)k^{2\beta+1}}   & 0 & 0 & \scriptscriptstyle d\zeta^\beta \left(1-\frac{\beta^2}{k\zeta}\right) \\
0 &\scriptscriptstyle \frac{\tilde{d}\Gamma(1+{\tilde{\beta}})e^{\imath2\pi{\tilde{\beta}}}\zeta^{-{\tilde{\beta}}-1}}{\tilde{r}\Gamma(-{\tilde{\beta}})l^{2{\tilde{\beta}}+1}} & \scriptscriptstyle \tilde{d}\zeta^{\tilde \beta} \left(1-\frac{\tilde \beta^2}{l\zeta}\right)&0 \\
0 &\scriptscriptstyle \frac{-\zeta^{-\tilde \beta} }{\tilde{d}}\left(1-\frac{\tilde \beta^2}{l\zeta}\right) &\scriptscriptstyle \frac{-\tilde{r}\Gamma(1-{\tilde{\beta}})e^{-\imath2\pi{\tilde{\beta}}}\zeta^{{\tilde{\beta}}-1}}{\tilde{d}\Gamma({\tilde{\beta}})l^{-2{\tilde{\beta}}+1}}&0 \\
\scriptscriptstyle \frac{-\zeta^{-\beta}}{d} \left(1-\frac{\beta^2}{k\zeta}\right) & 0 & 0 &\scriptscriptstyle\frac{-re^{-2\imath\pi\beta}\Gamma(1-\beta)\zeta^{{\beta}-1}}{d\Gamma(\beta)k^{-2\beta+1}}  \\
\end{pmatrix},
\end{align}
which is ascertained by noting the asymptotes of  $P,$ $Q$ for large  $|\zeta|$ :\begin{align}
Q^i_{\alpha}(\zeta)&=(-)^{i+1}e^{\imath \pi \alpha}\zeta^{\alpha-i }(1+O(\zeta^{-1}))+\frac{\Gamma(\alpha+1-i)e^{\imath 2\pi\alpha}}{\Gamma(-\alpha+i)}e^{-\zeta}\zeta^{-\alpha-1+i }(1+O(\zeta^{-1}))
\\P^i_\alpha(\zeta)&=(-)^{i}\zeta^{\alpha-i}(1+O(\zeta^{-1})).
\end{align}
The asymptotics of $Y^{(\rm mid)}_{II}$ in Eq. (\ref{Ymidin}) matches well that of   $Y^{(\rm out)}$ given in Eq. (\ref{Youtin}).  

The last requirement, namely that the asymptotics of $Y^{(\rm mid)}$ matches that of $Y^{\rm(out)}$  for all other regions (region $II$ being   just above considered), may be ascertained by noting that the jump matrices between regions $I$\ and $II$\ and between regions $II$ and $III$\ tend to unity, so the asymptotics of $Y^{(\rm mid)}$ in region $II$\ also determines the asymptotics throughout all the regions inside the unit circles (regions $I$-$III$), while $Y^{(\rm out)}$ also has the same asymptotics as the one  given in Eq. (\ref{Youtin}) inside the unit circle. To see the matching of asymptotes outside the unit circle, one may then first proceed to compute $Y^{\rm(mid)}_I$. It turns out that all exponential factors in $\zeta$ including those that  may be small in the unit circle but explode outside the unit circle vanish in region $I$. This can be seen by computing $Y_I^{\rm (mid)}$ explicitly by applying the appropriate jump matrix to  $Y_{II}^{\rm (mid)}$ :
\begin{align}
&Y_I^{(\rm mid)}=\\&=\nonumber\begin{pmatrix}\scriptscriptstyle\frac{d\Gamma(1+\beta)e^{\imath2\pi\beta}\psi^1_{\beta+1}(k\zeta)}{r\Gamma(-\beta)k^{2\beta+1}} & 0 & 0 & \scriptscriptstyle \frac{de^{\pi\imath\beta}\psi^1_{-\beta}(-k\zeta)}{k^\beta}  \\
0 &\scriptscriptstyle \frac{\tilde{d}\Gamma(1+{\tilde{\beta}})e^{\imath2\pi{\tilde{\beta}}}\psi^1_{\tilde \beta+1}(l\zeta)}{\tilde{r}\Gamma(-{\tilde{\beta}})l^{2{\tilde{\beta}}+1}} & \scriptscriptstyle \frac{\tilde de^{\pi\imath\tilde \beta}\psi^1_{-\tilde \beta}(-l\zeta)}{l^{\tilde{\beta}}} \\
0 &\scriptscriptstyle  \frac{-\psi^1_{\tilde \beta}(l\zeta)}{\tilde dl^{-\tilde{\beta}}} &\scriptscriptstyle \frac{-\tilde{r}e^{-\imath2\pi{\tilde{\beta}}}\Gamma(1-{\tilde{\beta}})\psi^1_{-\tilde \beta+1}(-l\zeta)}{\tilde{d}\Gamma({\tilde{\beta}})l^{-2{\tilde{\beta}}+1}} \\
\scriptscriptstyle\frac{-\psi^1_{\beta}(k\zeta)}{ dk^{-{\beta}}} & 0 & 0 &\scriptscriptstyle\frac{-re^{-2\imath\pi\beta}\Gamma(1-\beta)\psi^1_{-\beta+1}(-k\zeta)}{d\Gamma(\beta)k^{-2\beta+1}} \\
\end{pmatrix}.
\end{align}  
Indeed, all the elements of this matrix are proportional to confluent hypergeometric  which have a regular power expansion at large argument times $\zeta^{\pm\beta}$ or $\zeta^{\pm\tilde\beta}$.
This good behavior of $Y_{I}^{\rm(mid)}$ is shared by $Y_{VI}^{\rm(mid)}$ since the jump matrix between these two regions do not contain any exponential factors. Then, by an argument similar to the one above, which was applied to the regions interior to the unit circles, namely regions $I$-$III$, one can easily see that all regions outside the unit circle match asymptotically $Y^{\rm(out)}$.

This completes the proof that $Y_{II}^{\rm(mid)}$ given in Eq. (\ref{YIImid})
satisfies the requirements we have set for it, and we may turn to compute $Y_{}^{\rm(in)}$ . Before doing that we wish to extract the information that is useful for our final calculation of the entanglement  and the  negativity spectra. This is the following combination

\begin{align}
& \delta R^{(\rm mid/out)}\equiv Y^{\rm(mid)}_{II}Y^{\rm(out)-1}_{II}-\mathds{1}=-\frac{z_F^{\pm1}}{(z-z^{\pm1}_F)}\begin{pmatrix}   \frac{\beta ^2}{k} &0        &  0 &\dots\\
0  & \frac{\tilde \beta ^2}{l} &\dots  & 0  \\
      0   &  \dots & \frac{\tilde \beta ^2}{l}  &0  \\
 \dots &0 \frac{  }{}  &0  & \frac{\beta ^2}{k} 
\end{pmatrix} + O(1/\zeta^2),\label{deltaRmidout}
\end{align}
where the ellipsis denotes terms which will not serve our final calculation, but may be easily derived from the expressions above and the expansion here is in large $\zeta$.  

\subsection{Inner Region}In the inner region we wish to make the following transformation:
\begin{align}
&y(z)=Y^{(\rm in)} (z)O_R^{(\tau)}(z), \mbox{ for }|z|<1\\ & y(z)=Y^{(\rm in)}(z)O_L^{(\tau)}(z), \mbox{ for }|z|>1 , 
\end{align} 
where the matrices $O_{R/L}^{(\tau)}$ are given as follows: 
\begin{align}
&\label{OR}O^{(\tau)}_R=\begin{pmatrix}\sqrt{2}z^{-(k+m)}  & \frac{-\tau^2z^{-(k+m)}}{\sqrt{2}} &  &  \\
\tau\sqrt{2} z^{-l}  & \frac{z^{-l}}{\tau\sqrt{2}}  &  &  \\
 &  & \frac{\tau}{\sqrt{2}}  & \frac{1}{\tau\sqrt{2}}  \\
 &  & \frac{-\tau^2z^{-m}}{\sqrt{2}}  & \frac{z^{-m}}{\sqrt{2}}  \\
\end{pmatrix}  ,\\& \label{OL} O^{(\tau)}_L=\begin{pmatrix}\sqrt{2} z^{-m} & \frac{-\tau^2z^{-m}}{\sqrt{2}}  &
\frac{-\tau^2\lambda z^{-m}}{\sqrt{2}}  & \frac{\lambda z^{-m}}{\sqrt{2}}  \\ \tau \sqrt{2}  & \frac{1}{\tau\sqrt{2}}  & \frac{\tau \lambda  }{\sqrt{2}}  & \frac{\lambda }{\tau \sqrt{2}}  \\
 &  & \frac{\tau z^{-l}}{\sqrt{2} }  & \frac{ z^{-l}}{\tau\sqrt{2} }  \\
 &  & \frac{-\tau^2z^{-(k+m)}}{\sqrt{2}}  & \frac{z^{-(k+m)}}{\sqrt{2} }  \\
\end{pmatrix}.\end{align}

The matrix $y$ can easily be seen to obey a Riemann-Hilbert problem:\begin{align}
y(e^{\imath x-0^+})=y(e^{\imath x+0^+})v\label{oldyeq},
\end{align}
 for $x$ real and where the jump matrix $v$ is related to the old jump martix $V^{}$ through
\begin{align}V^{}O^{(\tau)}_R=O^{(\tau)}_L v .\end{align}
Explicitly, given the choice in Eqs. (\ref{OR},\ref{OL})  for the matrices $O^{(\tau)}_R $ and $O^{(\tau)}_L$, the jump matrix $v$ is given by:
\begin{align}
v\label{littlev}=\begin{pmatrix}1 &  &  & f  \\
 & 1 &  &  \\
 &  & 1 &  \\
 &  &  & 1 \\
\end{pmatrix}.
\end{align}
If we now compute the monodromy around the origin, the appropriate monodromy matrix is given by:
\begin{align}
\begin{pmatrix}1 &  &  & f_i-f_o  \\
 & 1 &  &  \\
 &  & 1 &  \\
 &  &  & 1 \\
\end{pmatrix}=
\begin{pmatrix}1 &  &  & 2\imath r\sin(\pi\beta)  \\
 & 1 &  &  \\
 &  & 1 &  \\
 &  &  & 1 \\
\end{pmatrix}=
\begin{pmatrix}1 &  &  & 2\imath \tau^2\tilde r\sin(\pi\tilde \beta)  \\
 & 1 &  &  \\
 &  & 1 &  \\
 &  &  & 1 \\
\end{pmatrix}.   
\end{align}

On the other hand, given $j$, the $j$-th row of the matrix $y_{II}$, that we shall denote by $\bm v^{(j)}_y$, is given by the $j$-th row of $Y^{\rm (in)}_{I},$ denoted by, $\bm v^{(j)}_Y,$ as follows 
\begin{align}
\boldsymbol{v}^{(j)}_Y=\boldsymbol{v}^{(j)}_y\cdot
\begin{pmatrix}\frac{e^{m\zeta}}{2\sqrt{2}}\left(\frac{ f}{re^{\imath \pi\beta}}+1\right)  & \frac{1 }{2\tau \sqrt{2}} & 0 \frac{- e^{m\zeta}f}{2\sqrt{2} re^{\imath \pi\beta}} \\
\frac{e^{m\zeta}\tau^2}{\sqrt{2}}\left(\frac{ f}{re^{\imath \pi\beta}}-1\right) & \frac{\tau }{\sqrt{2}} & 0 & \frac{- e^{m\zeta}f\tau^2}{\sqrt{2} re^{\imath \pi\beta}} \\
\frac{\tau^2e^{m\zeta}}{\sqrt{2} re^{\imath \pi\beta}} & \frac{-1}{\sqrt{2}    \tau\tilde  r^{}e^{\imath \pi\tilde \beta^{}}} & \frac{1 }{\sqrt{2}\tau} & \frac{ -\tau^2e^{m\zeta} }{\sqrt{2}} \\
\frac{-e^{m\zeta}}{\sqrt{2} re^{\imath \pi\beta}} & \frac{-\tau}{\sqrt{2}    \tilde r^{}e^{\imath \pi\tilde \beta^{}}} & \frac{\tau }{\sqrt{2}} & \frac{ e^{m\zeta} }{\sqrt{2}} \\
\end{pmatrix}.
\end{align}
In order for all factors of the form $e^{\pm m\zeta}$ to disappear from $\bm{v}_Y^{(j)}$ which is required in order to make them disappear also from $Y_{VI}^{\rm(in)},$ since $V_{I,VI}$ does not have exponential factors, we must demand that $\bm v^{(j)}_y$ is of the form:
\begin{align}
\bm v^{(j)}_y=&\left(\tau^2F(\zeta) -\tau^2H(\zeta)e^{-m\zeta},\frac{F(\zeta)}{2} +\frac{H(\zeta)e^{-m\zeta}}{2}\right.,\\&\left.,\tau^2G(\zeta)-\frac{F(\zeta)  f}{2}-\tau^2
J(\zeta)e^{-m\zeta}, G(\zeta)+\frac{F(\zeta) f\tau^2}{2}+J(\zeta)e^{-m\zeta}\right),
\end{align} 
where $F$, $G$, $H$ and $J$\  remain well-bounded in the inner region. Since the first, second and third elements of $\bm v_y$ must be analytical by the Riemann-Hilbert problem, one concludes immediately that $F(\zeta)$ and $H(\zeta)$ must be analytical as well. 

We shall now use the fact that the following function is analytical around the origin having no jumps or singularities: \begin{align}
&P^{i}_{\tilde \beta}(l\zeta)+s_{\tilde\beta} Q^i_{\tilde\beta}(l\zeta)\Gamma_0(m\zeta)\\
&P^1_{ \beta}(k\zeta)+s_\beta Q^i_{\beta}(k\zeta)\Gamma_0(-m\zeta),
\end{align}
where the symbol $s_\alpha$ denotes:

\begin{align}
s_\alpha =\frac{\sin \pi\alpha}{\pi} 
\end{align}and  $\Gamma_0(\zeta)$ is the incomplete Gamma function, which is denoted usually by $\Gamma(0,\zeta), $ here  a different notation being employed for the sake of brevity. The analytical properties around the origin of the combinations above can be ascertained by considering that the monodromy around zero vanishes and by noting that the function is not singular at the origin.  Here one is aided by the following monodromy of $\Gamma_0(\zeta)$, given by $\Gamma_0(e^{2\pi\imath}\zeta)=\Gamma_0(\zeta)+2\pi\imath$.

  The function $\Gamma_0(\pm m\zeta)$ may be considered as the function $e^{\pm m\zeta}\Gamma_0(\pm m\zeta),$ which has a regular expansion at $\zeta\to\infty$ times $e^{\mp m\zeta}$, these regular expansions are of order $\zeta^{-1}$. With this we may then construct solution for the first  and fourth row of $y_{II}$:
\begin{align}
&\bm v^{(k)}_y\label{vy14}=C_k\left(\frac{2}{r}Q^{i_k}_\beta(k\zeta) ,-\frac{\tau^2}{r  }Q^{i_k}_\beta(k\zeta),-\tau^2\left(P^{i_k}_\beta(k\zeta)+s_\beta Q^{i_k}_\beta(k\zeta)\Gamma_0(-m\zeta)\right)\nonumber, \right.\\&\left.,P^{i_k}_\beta(k\zeta)-s_\beta Q^{i_k}_\beta(k\zeta)\Gamma_0(-m\zeta)\right).\nonumber \end{align}
 where $k\in\{1,4\}$
 \begin{align}
 &C_1=\frac{    de^{-m\zeta}}{ \sqrt2k^\beta}, \quad C_4=\frac{   re^{-m\zeta}e^{-\imath2\pi\beta}\Gamma(1-\beta)}{d\sqrt{2}\Gamma(\beta)k^{-\beta}}
\\
& i_1=0,\quad  \quad i_4=1, \quad  \end{align}
For the second and third row we offer the following solution:\begin{align}
&\bm v^{(k)}_y\label{vy1}=C_k\left(\frac{2\tau^2}{\tilde r}Q^{i_k}_{\tilde\beta}(l\zeta) ,\frac{1}{\tilde r}Q^{i_k}_{\tilde\beta}(l\zeta),\tau^2\left(P^{i_k}_{\tilde{\beta}}(l\zeta)+s_{\tilde{\beta}} Q^{i_k}_{\tilde\beta}(l\zeta)\Gamma_0(m\zeta)\right)\nonumber-\frac{fQ^{i_k}_{\tilde\beta}(l\zeta)}{\tilde r},\right.\\&\left., P^{i_k}_{\tilde{\beta}}(l\zeta)-s_{\tilde{\beta}} Q^{i_k}_{\tilde\beta}(l\zeta)\Gamma_0(m\zeta)+\frac{fQ^{i_k}_{\tilde\beta}(l\zeta)}{\tilde r}\right). \end{align}
where $k\in\{2,3\}$ and
\begin{align}
 &C_2=\frac{    \tilde d}{ \sqrt2\tau l^{\tilde \beta}}, \quad C_3=\frac{   \tilde re^{-\imath2\pi\tilde \beta}\Gamma(1-\tilde \beta)}{\tilde d\sqrt{2}\tau^2\Gamma(\tilde \beta)l^{-\tilde \beta}}
\\
& i_2=0,\quad  \quad i_3=1, \quad  \end{align}

To ascertain that the rows  above of $\bm v_y^{(j)}$ are indeed correct, we now derive from it the matrix $Y^{\rm(in)}_{II}$  written here up to exponentially small terms in $m|\zeta|$ as $m|\zeta|\to\infty$  (which exist in this region but not, e.g., in region $I$ by construction)
:     

\begin{align}
&Y^{(\rm in)}_{II}=\\&=\nonumber
\begin{pmatrix}[1.7] \scriptscriptstyle  \frac{d}{r k^\beta}e^{\zeta k}Q^0_\beta(k\zeta) & 0 &\scriptscriptstyle \frac{-\tau ds_\beta  Q^0_\beta(k\zeta)e^{-m\zeta}\Gamma_0(-m\zeta)}{ k^\beta} & \scriptscriptstyle \frac{d}{k^\beta} P^0_\beta(k\zeta) \\
0 & \scriptscriptstyle \frac{\tilde d}{\tilde rl^{\tilde{\beta}}}e^{\zeta l} Q^0_{\tilde\beta}(l\zeta) &\scriptscriptstyle\frac{\tilde d }{l^{\tilde{\beta}}}P^0_{\tilde\beta}(l\zeta)  & \scriptscriptstyle\,\frac{-\tilde{d} s_{\tilde\beta}Q^0_{\tilde\beta}(l\zeta)e^{m\zeta}\Gamma_0(m\zeta)}{ \tau l^{\tilde{\beta}}   }   \\
        0 &  \scriptscriptstyle\frac{-\Gamma(1-{\tilde{\beta}})e^{\zeta l}Q^{1}_{ {\tilde{\beta}}}(l\zeta)}{\tilde{d}e^{\imath2\pi{\tilde{\beta}}}\Gamma({\tilde{\beta}}) l^{-\tilde{\beta}}} &\scriptscriptstyle\frac{- \tilde r\Gamma(1-{\tilde{\beta}})P^1_{\tilde \beta}(l\zeta)}{\tilde{d} e^{\imath2\pi{\tilde{\beta}}}\Gamma({\tilde{\beta}})l^{-\tilde{\beta}}}   &\scriptscriptstyle \,\frac{\tilde r \Gamma^{}(1-{\tilde{\beta}}) s_{\tilde\beta}Q^1_{ {\tilde{\beta}}}(l\zeta)e^{m\zeta}\Gamma_0(m\zeta)}{\tilde{d}\tau e^{\imath2\pi{\tilde{\beta}}}\Gamma({\tilde{\beta}}) l^{-\tilde{\beta}}   }  \\
\scriptscriptstyle \frac{-\Gamma(1-\beta)e^{\zeta k}Q^1_\beta(k\zeta)}{de^{\imath2\pi\beta}\Gamma(\beta)k^{-\beta}} & 0 & \scriptscriptstyle  \frac{\tau r\Gamma(1-\beta)e^{\zeta k}s_\beta Q^1_\beta(k\zeta)e^{-m\zeta}\Gamma_0(-m\zeta)}{de^{\imath2\pi\beta}\Gamma(\beta)k^{-\beta}}  & \scriptscriptstyle \frac{-r\Gamma(1-\beta)e^{\zeta k}P^1_\beta(k\zeta)}{de^{\imath2\pi\beta}\Gamma(\beta)k^{-\beta}} 
\end{pmatrix} .
\end{align}

Now  using:
\begin{align}
Q^{0}_{\alpha}(\zeta)\overset{\zeta\to0}\to  \frac{-e^{\imath \pi\alpha}}{\Gamma(-\alpha)s_\alpha},\quad Q^{1}_{\alpha}(\zeta)\overset{\zeta\to0}\to  \frac{-e^{\imath \pi\alpha}}{\Gamma(1-\alpha)s_\alpha} \quad 
\end{align}we have: 
\begin{align} Y^{(\rm in)}_{II}-Y^{(\rm mid)}_{II}\overset{k\zeta,l\zeta\to0}\longrightarrow\begin{pmatrix}   0 & 0 &\scriptscriptstyle \frac{\tau de^{\imath \pi\beta}  e^{-m\zeta}\Gamma_0(-m\zeta)}{ \Gamma(-\beta)k^\beta} &  0 \\
0 & 0 &0  & \scriptscriptstyle\,\frac{\tilde{d}e^{\imath \pi\tilde \beta} e^{m\zeta}\Gamma_0(m\zeta)}{ \tau\Gamma(-\tilde \beta) l^{\tilde{\beta}}   }   \\
        0 &  0 & 0  &\scriptscriptstyle \,\frac{-\tilde re^{-\imath\pi{\tilde{\beta}}} e^{m\zeta}\Gamma_0(m\zeta)}{\tilde{d}\tau\Gamma({\tilde{\beta}}) l^{-\tilde{\beta}}   }  \\
 0 & 0 &\scriptscriptstyle \frac{-\tau re^{-\imath \pi\beta}e^{\zeta k} e^{-m\zeta}\Gamma_0(-m\zeta)}{d\Gamma(\beta)k^{-\beta}}  & 0 \\
\end{pmatrix} 
\end{align}
We shall want to compute $Y^{(\rm in)}_{II} Y^{(\rm mid)-1}$, which will be important in the following. To this aim we first write:  
\begin{align}
& Y^{\rm(mid)-1}_{II}(\zeta)=\\&\nonumber\begin{pmatrix}\scriptscriptstyle\frac{  -re^{-\imath2\pi\beta}\Gamma(1-\beta)P^1_{ \beta}(k\zeta)}{d\Gamma(\beta)k^{-\beta}}   & 0 & 0 & \scriptscriptstyle \frac{-dP^0_{ \beta}(k\zeta)}{k^\beta}  \\
0 &\scriptscriptstyle\frac{-\tilde{r} e^{-\imath2\pi{\tilde{\beta}}}\Gamma(1-{\tilde{\beta}})P^1_{ \tilde \beta}(l\zeta)}{\tilde{d}\Gamma({\tilde{\beta}})l^{-\tilde{\beta}}}   & \scriptscriptstyle \frac{-\tilde dP^0_{ \tilde \beta}(l\zeta)}{l^{\tilde{\beta}}} \\
0 &\scriptscriptstyle  \frac{e^{-\imath2\pi{\tilde{\beta}}}\Gamma(1-{\tilde{\beta}})e^{\zeta l}Q^1_{ \tilde \beta}(l\zeta)}{\tilde{d}\Gamma({\tilde{\beta}})l^{-\tilde{\beta}}}  &\scriptscriptstyle \frac{\tilde de^{\zeta l}Q^0_{ \tilde \beta}(l\zeta)}{\tilde rl^{\tilde{\beta}}} \\
\scriptscriptstyle\frac{-e^{-\imath2\pi\beta}\Gamma(1-\beta)e^{\zeta k}Q^1_{ \beta}(k\zeta)}{d\Gamma(\beta)k^{-\beta}} & 0 & 0 &\scriptscriptstyle \frac{de^{\zeta k} Q^0_{ \beta}(k\zeta)}{r k^\beta}  \\
\end{pmatrix},
\end{align}
from which the following may easily be computed:

\begin{align}
&  \delta R^{(\rm in/mid)}\equiv Y^{(\rm in)}_{II} Y^{(\rm mid)-1}_{II} -\mathds{1}\overset{k\zeta,l\zeta\to0}{\longrightarrow}\frac{z^{\pm1}_F}{m(z-z^{\pm1}_F)}\label{deltaRinmid}\times\\&\nonumber\times\begin{pmatrix}   0 &\scriptscriptstyle \frac{-\tau de^{\imath \pi(\beta-\tilde{\beta})}l^{\tilde{\beta}}  }{ \tilde{d}\Gamma(\tilde\beta)\Gamma(-\beta)s_{\tilde\beta}k^\beta}   &\scriptscriptstyle \frac{\tau\tilde{d}de^{\imath \pi(\beta+\tilde\beta)}  }{ \tilde{r}\Gamma(-\tilde{\beta})\Gamma(-\beta)s_{\tilde{\beta}}k^\beta l^{\tilde\beta}}   &  0 \\
\scriptscriptstyle \frac{\tilde {d}e^{\imath\pi(\tilde \beta-\beta)} k^{\beta}}{\tau d\Gamma(-\tilde \beta)\Gamma(\beta)s_\beta  l^{\tilde{\beta}} } & 0 &0  & \scriptscriptstyle\,\scriptscriptstyle\frac{-d\tilde{d}e^{\imath \pi(\tilde \beta+\beta)} }{ \tau r\Gamma(-\tilde \beta)\Gamma(-\beta)s_\beta k^\beta l^{\tilde{\beta}}   }  \\
      \scriptscriptstyle \frac{- \tilde r e^{-\imath\pi(\beta+\tilde{\beta})} l^{\tilde{\beta}}k^{\beta}}{\tau\tilde d d\Gamma(\tilde\beta)\Gamma(\beta)s_\beta }  &  0 & 0  &\scriptscriptstyle \frac{ d\tilde re^{\imath\pi(\beta-{\tilde{\beta}}) l^{\tilde{\beta}}} }{\tilde{d}r\tau\Gamma({\tilde{\beta}})\Gamma(-\beta)s_\beta k^\beta   }  \\
 0 &\scriptscriptstyle \frac{  }{} \frac{\tau re^{-\imath \pi(\beta+\tilde\beta)}l^{\tilde{\beta}}k^{\beta} }{d\tilde{d}\Gamma({\tilde{\beta}})\Gamma(\beta)s_{\tilde\beta}} &\scriptscriptstyle \frac{-\tilde dr\tau e^{\imath \pi(\tilde{\beta}-\beta)}k^{\beta} }{d\tilde r\Gamma(\beta)\Gamma(-\tilde{\beta})l^{\tilde{\beta}}s_{\tilde{\beta}}} &         0 
\end{pmatrix} 
+O(1/\zeta^2).\end{align}
With this in hand we may proceed to discuss how to combine the middle, inner and outside regions.

\subsection{Combining  the three Regions}

We have solved the Riemann-Hilbert problem in the outer region middle and  inner regions separately, making sure that the solutions match to leading order on the boundary between the two regions (the gray circles in Fig. \ref{Decomposition}). In order to find a global solution one may write and solve a Riemann-Hilbert problem for the discongruity, following \cite{Deift:Its:Krasovksy:Toeplitz:Hankel}.  Indeed, one may define and then solve the Riemann-Hilbert problem for the object:
\begin{align}
\mathcal{R}(z)=\begin{cases}Y^{}(z)Y^{\rm (out)-1}(z) & z\mbox{  in outer Region} \\
Y^{}(z)Y^{\rm (mid)-1}(z) & z\mbox{  in middle Region}\\
Y^{}(z)Y^{\rm (in)-1} (z) & z\mbox{  in inner Region} \\
\end{cases}\label{ysandrs}
\end{align}
The jump matrix for this  problem is  exponentially small on the unit circle while on the border between the inner and outer regions it is given by:
\begin{align}
&V^{(\rm mid/out)}=\mathds{1}+\delta R^{(\rm mid/out)},\quad V^{(\rm in/mid)}=\mathds{1}+\delta R^{(\rm in/mid)},
\end{align}
where $\delta R$\ is given in Eqs. (\ref{deltaRinmid},\ref{deltaRmidout}).
Crucially,  the jump matrices of the new Riemann Hilbert problem are always close to the identity so it is easy to offer a solution for the Hilbert-Riemann problem  to $\mathcal{R}$ any given order. The leading terms of the solution at the outer region is given as follows:
\begin{align}
&\mathcal{R}^{\rm(out)}(z)\label{Rout}
=\mathds{1}+\sum_{I\in\{L,R\}}\left(\delta R^{\rm (in/mid)}_I(z)+ R^{\rm (mid/out)}_I(z)\right)+\\&+\delta R^{\rm (in/mid)}_R(z_F)\delta R^{\rm (in/mid)}_L(z)+\delta R^{\rm (in/mid)}_L(z^{-1}_F)\delta R^{\rm (in/mid)}_R(z)+\dots\nonumber\end{align}
where the index $I$\ takes the values $R$ or $L$ depending the relevant $\delta R$ relates to the expansion near $z^{-1}_F$ or $z_F^{}$, respectively. We have not written the second order in $R^{\rm (mid/out)}$, since these are sub-leading, while the second order in $R^{\rm (mid/out)}$ was written since this is the first order containing diagonal elements (in contrast to  $R^{\rm (mid/out)}$ contains diagonal terms already in the first order).

 We are interested in the diagonal elements of $\mathcal{R}^{\rm (out)}(0)$  as these element  features in the final result for both negativity and entanglement. Indeed, by combining Eqs. (\ref{DetsAndChis}, \ref{TsandChis12}, \ref{ysandrs}), we have:
\begin{align}
&\mathcal{R}_{22}^{\rm (out)}(0)Y^{(\rm out)}_{23}(0)=\frac{-D_{k,l,n}}{D_{k,l-1,n}},\quad  \mathcal{R}_{11}^{\rm (out)}(0)Y^{(\rm out)}_{14}(0)=\frac{-D_{k,l,n}}{D_{k,l-1,n}}.\quad\label{Go}
\end{align} It is possible to obtain a result for $\mathcal{R}^{(\rm out)}_{1}$, $ \mathcal{R}^{(\rm out)}_{22}$  by using Eqs. (\ref{Rout},\ref{deltaRinmid},\ref{deltaRmidout}) :
\begin{align}
&\label{Rexplicit22}\mathcal{R}_{22}^{\rm (out)}(0;\lambda)=1-\frac{\tilde \beta^{2}}{l}+\frac{1}{m^22\imath\sin p_F s_\beta  s_{\tilde{\beta}}}\left[\frac{e^{-\imath p_F}(2k\sin p_F)^{2\beta}}{\Gamma^2(-\tilde \beta)\Gamma^2(\beta)(2l\sin p_F)^{2\tilde \beta}} \right.+\\&+\left.\frac{e^{\imath p_F}(2k\sin p_F)^{2\beta} (2l\sin p_F)^{2\tilde \beta} }{ \Gamma^2(\tilde \beta)\Gamma^2(\beta)    }\right]+(\beta,\tilde \beta,p_F\to-\beta,-\tilde \beta,-p_F),\nonumber\\
&\label{Rexplicit11}\mathcal{R}_{11}^{\rm (out)}(0;\lambda)=1-\frac{\beta^2}{k}+\frac{1}{m^22\imath\sin p_F s_\beta  s_{\tilde{\beta}}}\left[\frac{e^{\imath p_F}(2k\sin p_F)^{2\beta}}{\Gamma^2(-\tilde \beta)\Gamma^2(\beta)(2l\sin p_F)^{2\tilde \beta}} \right.+\\&+\left.\frac{e^{\imath p_F}(2k\sin p_F)^{2\beta} (2l\sin p_F)^{2\tilde \beta} }{ \Gamma^2(\tilde \beta)\Gamma^2(\beta)    }\right]+(\beta,\tilde \beta,p_F\to-\beta,-\tilde \beta,-p_F),\nonumber
\end{align}
where $+(\beta,\tilde \beta\to-\beta,-\tilde \beta)$     denotes adding the same expressions with $\beta$ and $\tilde \beta^{}$ are reversed in sign, and we have included $\lambda$ as an explicit argument of $ \mathcal{R}^{\rm (out)}_{22}$  . We have assumed here $F_\pm^{(\sigma)}(z_F)=F_\pm^{(\sigma)}(z^{-1}_F).$ In deriving Eqs.  (\ref{Rexplicit22},\ref{Rexplicit11}), we have used Eqs. (\ref{ddef}). We may compute the logarithmic derivative of the ratio of determinants to obtain the change of the spectrum of eigenvalues as, e.g.,  $l$ is changed as the jump discontinuity of the object thus derived:\begin{align}
&d\log\frac{D_{k-1,l,n}}{D_{k-1,l-1,n}}=d \log \ \left[\mathcal{R}_{22}^{\rm (out)}(0)Y^{(\rm out)}_{23}(0)\right]=d \log\left[ \mathcal{R}_{22}^{\rm (out)}(0)\nonumber z_F^{ 2\tilde{\beta}}e^{-\imath \pi\tilde {\beta}}\tilde F^{}_+(0)\right]=\\
&=d\lambda \oint \frac{1}{\lambda+\tau^2f(\theta)} \frac{d\theta}{2\pi }+d\log \mathcal{R}_{22}^{\rm (out)}(0),\label{dlogDet}
\end{align}
where we have used  Eqs.  (\ref{SzegoComesfrom},\ref{Youtin}) to derive the last line. If $k$ is changed rather than $l$ then one to only replace $d\log \mathcal{R}_{22}^{\rm (out)}(0)$ by $d\log \mathcal{R}_{11}^{\rm (out)}(0)$  and $\lambda+\tau^2f(\theta)$ by $\lambda+f(\theta)$ in the denominator of the integrand in the last line of Eq. (\ref{dlogDet}). 
 
Eq. (\ref{dlogDet}) has a pole singularity on eigenvalues of the matrices $D_{k-1,l,n}$ and  $D_{k-1,l-1,n}$. The change of the spectral density,  $\Delta d\omega^{(\tau)}$,    is then given by  $\frac{1}{2\pi\imath}$ times the jump discontinuity of  the expression on the right hand side of Eq. (\ref{dlogDet}) , namely we have :
\begin{align} 
\Delta d\omega^{(\tau)}\left(\lambda\right)=\frac{1}{2\pi}|d\theta(\pm\lambda)|+\frac{1}{2\pi\imath}d\log \frac{\mathcal{R}^{(\rm out)}_{jj}(0;\lambda+\imath 0^+)}{
\mathcal{R}_{jj}^{(\rm out )}(0;\lambda-\imath 0^+)}\label{FinalResult},
\end{align} 
where $\theta(f)$ is the inverse function to $f(\theta)$ and $j$ takes the value $1$ if $\Delta$ implies changing $k$ otherwise, if $\Delta$ implies a change in $l,$ then $j$ takes the value $2$. Thus we have the following expressions as our final result:\begin{align} 
&\Delta_l d\omega^{(\tau)}\left(\lambda\right)=\frac{1}{2\pi}|d\theta(-\tau^2\lambda)|+\frac{1}{\pi}\Im\left[-\frac{\tilde \beta^{2}}{l}+\frac{1}{m^22\imath\sin p_F s_\beta  s_{\tilde{\beta}}}\left(\frac{e^{-\imath p_F}(2k\sin p_F)^{2\beta}}{\Gamma^2(-\tilde \beta)\Gamma^2(\beta)(2l\sin p_F)^{2\tilde \beta}} \right.\right.\nonumber+\\&+\left.\left.\frac{e^{\imath p_F}(2k\sin p_F)^{2\beta} (2l\sin p_F)^{2\tilde \beta} }{ \Gamma^2(\tilde \beta)\Gamma^2(\beta)    }\right)+(\beta,\tilde \beta,p_F\to-\beta,-\tilde \beta,-p_F)\right]\label{finalfinal1},\\
&\Delta_k d\omega^{(\tau)}\left(\lambda\right)=\frac{1}{2\pi}|d\theta(-\lambda)|+\frac{1}{\pi}\Im\left[-\frac{\beta^2}{k}+\frac{1}{m^22\imath\sin p_F s_\beta  s_{\tilde{\beta}}}\left(\frac{e^{\imath p_F}(2k\sin p_F)^{2\beta}}{\Gamma^2(-\tilde \beta)\Gamma^2(\beta)(2l\sin p_F)^{2\tilde \beta}} \right.\right.\nonumber+\\&+\left.\left.\frac{e^{\imath p_F}(2k\sin p_F)^{2\beta} (2l\sin p_F)^{2\tilde \beta} }{ \Gamma^2(\tilde \beta)\Gamma^2(\beta)    }\right)+(\beta,\tilde \beta,p_F\to-\beta,-\tilde \beta,-p_F)\right].\label{finalfinal2}
\end{align} 

{The part of these equation featuring the inverse function, $\theta(f),$ of the Fermi occupation function, $f(\theta)$, gives rise to the regular entropy associated with a varying occupation of the Fermions. The part terms $\frac{\beta^2}{k},$ $\frac{\tilde{\beta}^2}{l}$ already appear in the computation of the entanglement  entropy of a single interval in the works of Ref. \cite{Jin:Korepin:Entanglement:Entropy:Fermions}, and thus may also be considered as a known result. The rest of the expression then represents the new result. We remind the reader that the variables $\beta$ and $\tilde{\beta}$ encode the jump discontinuity of the fermion occupation number about the Fermi point according to  Eq. (\ref{rbeta}). It is through these variables that one can distinguish between open and closed systems in our case. Note though that if further singularities, in addition to the jump at the Fermi points, occur in the fermion occupation number due to the opening of the system (which we have assumed here does not happen), a different Riemann-Hilbert analysis is required to deal with these singularities and thus Eqs. (\ref{finalfinal1}-\ref{finalfinal2}) no longer hold. On the other hand, the mapping of the problem to a Riemann-Hilbert problem, which we have described here, does continue to hold.}

Averages such as the entropy, $S$, can be computed by integrating with the measure $d\omega$. For example, denoting by $S^{(\tau)}$ the entropy of the two intervals, with or without a partial transpose of one of the regions, one has:
\begin{align}
\Delta S^{(\tau)}=\int \left(\frac{1+\lambda}{2}\log\frac{1+\lambda}{2}+\frac{1-\lambda}{2}\log\frac{1-\lambda}{2}\right)\Delta d\omega^{(\tau)}(\lambda),\label{entropychange}
\end{align} 
where again  $\Delta$ denotes the change in the quantity that follows (here the entropy) when one increases $l$ or $k$ by $1$.

\begin{figure}[h!!!]
\begin{center}
\includegraphics[width=10cm]{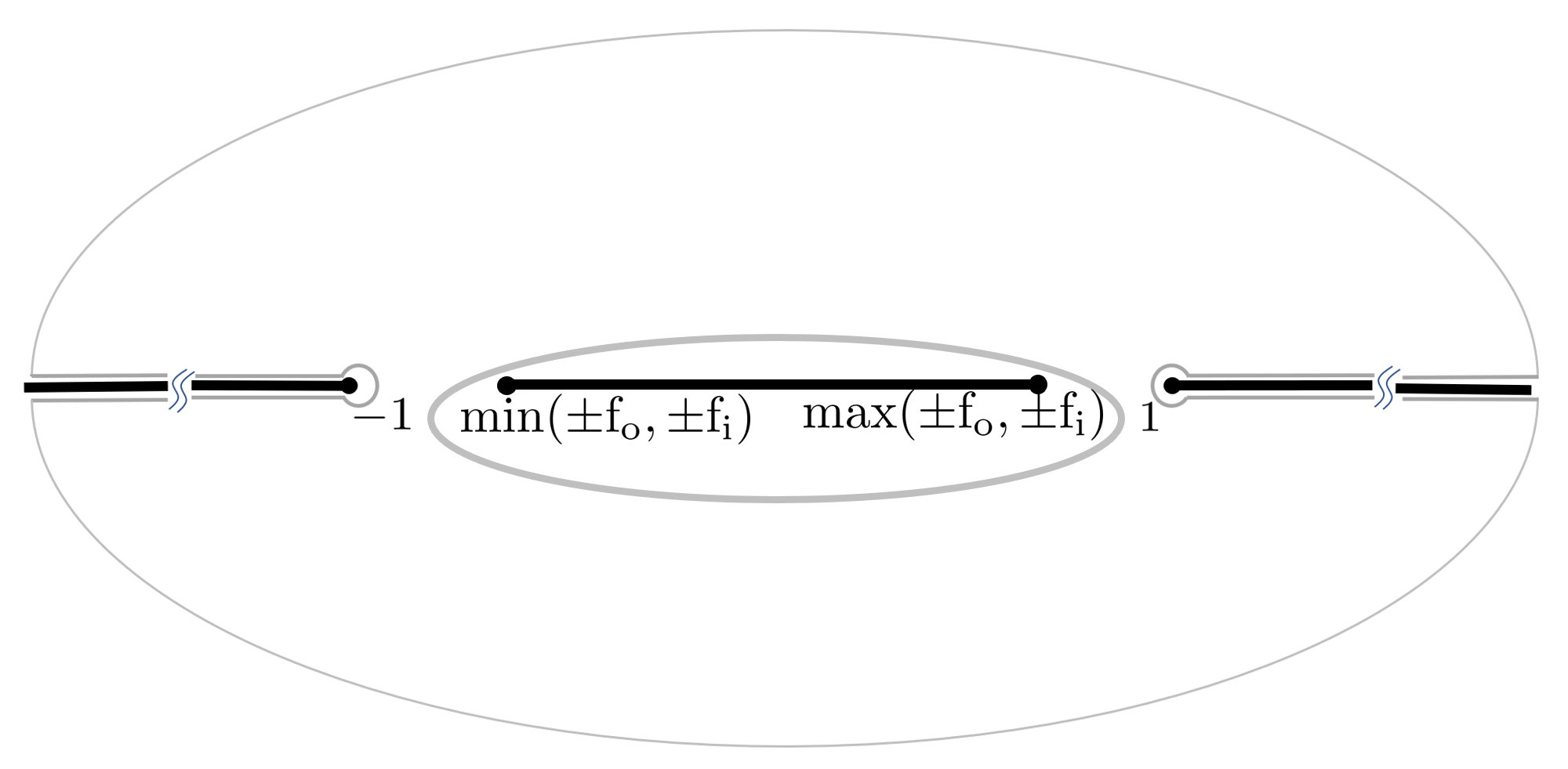}
\caption{
The contour of integration of Eq. (\ref{entropychange}). Originally drawn as the dashed line surrounding the interval $[\min (\pm f_o,\pm f_i),\max (\pm f_o,\pm f_i)]$ which necessarily contains the branch cuts of $\beta$ and $\tilde \beta$, is deformed to surround the branch cuts of $S(\lambda),$ where $S(\lambda)$ is the integrand in (\ref{entropychange}),  and a to include the large arcs as described by the gray line. The functions $\beta$ and  $\tilde\beta$ have small real parts on the gray line.  \label{contourdeformation}} \end{center}
\end{figure}

Another comment is in order. Strictly speaking, we have shown that the result of Eq. (\ref{FinalResult}) holds only in the case where the real parts of $\beta$ and $\tilde \beta^{}$ are small. This usually is no restriction, since when computing averages one is often able to deform the contour of integration to a region where this condition holds, to justify the use of the expression we have given. For example when computing the entropy, one has to compute the integral in Eq.(\ref{entropychange}).
This integral  can be performed in the manner shown in Fig. \ref{contourdeformation} such that indeed $\beta$ and $\tilde \beta$ are small, this justifies the use of the expressions we have for $d\omega^{(\tau)}$, obtained by using Eq. (\ref{FinalResult}). 

\section{Conclusion}

We have given a result for the change in the entanglement  and negativity  spectrum when the size of one of the intervals is changed in the case where the size of the interval and the distance between the intervals are all much larger than $1$. The result is given in Eq. (\ref{finalfinal1},\ref{finalfinal2}), which requires for its application the  definition in Eqs. (\ref{rbeta}) and knowledge of the fact that  $\tau$ is to be taken equal to $1$ for the computation of the entanglement entropy and $\imath$ if  negativity is computed. For example, to compute the entropy change under the change of the size of  the intervals one may use Eq. (\ref{entropychange}). The change in $l$ represents increasing the size of one of the intervals without changing the distance between them, while a change in $k$ represents the increasing the size of one of the intervals on expense of the distance between them.

It would be interesting to apply the method shown here in order to compute entanglement entropy and the negativity spectra in different physical situations, the case where the distance between the intervals is much larger than their size, follows straightforwardly by applying the results here to compute physical quantities, while the more general case {where the interval separation is arbitrary} requires a full solution of the Riemann-Hilbert problem beyond the limit found here. We plan to return to these questions in a future publication.

\section{Acknowledgement}
 E.B. and A. B. would like to acknowledge money from ISF grant number 1466/15.

\end{document}